\documentclass[letterpaper,12pt]{article}
\usepackage{setspace}
\usepackage[left=1in, top=1.5in, right=1in, bottom=1.5in]{geometry}
\usepackage[T1]{fontenc}
\usepackage[charter]{mathdesign}
\usepackage{amsmath}
\usepackage{amsthm}
\usepackage{appendix}
\usepackage{booktabs}
\usepackage{csquotes}
\usepackage{enumerate}
\usepackage{float}
\usepackage{graphicx}
\usepackage{placeins}
\usepackage{bm}
\usepackage{dsfont} %
\usepackage{ulem} %
\usepackage{proof-at-the-end}
\usepackage{makecell} %
\usepackage{xcolor}
\usepackage{subcaption}
\usepackage{textcomp}
\usepackage[backend=biber,
	authordate,
	url=false,
	doi=false,
	isbn=false,
	eprint=false]{biblatex-chicago}
\makeatletter
\newenvironment{subtheorem}[1]{%
  \def\subtheoremcounter{#1}%
  \refstepcounter{#1}%
  \protected@edef\theparentnumber{\csname the#1\endcsname}%
  \setcounter{parentnumber}{\value{#1}}%
  \setcounter{#1}{0}%
  \expandafter\def\csname the#1\endcsname{\theparentnumber-\arabic{#1}}%
  \expandafter\def\csname theH#1\endcsname{thm.\theparentnumber-\arabic{#1}}%
  \unskip\ignorespaces
}{%
  \setcounter{\subtheoremcounter}{\value{parentnumber}}%
  \ignorespacesafterend
}
\makeatother
\newcounter{parentnumber}

\theoremstyle{plain}%

\newtheorem*{LM*}{Lemma}

\newtheorem*{CR*}{Corollary}
\newtheorem{HP}{Hypothesis}
\newtheorem{RS}{Result}

\newtheorem*{OBS}{Observation}

\newtheorem*{thm*}{Theorem}

\newtheorem*{lemma*}{Lemma}

\newtheorem*{proposition*}{Proposition}

\newtheorem*{corollary*}{Corollary}

\newtheorem*{conjecture*}{Conjecture}

\newcounter{HPc}

\theoremstyle{definition}
\newtheorem{DF}{Definition}
\newtheorem{EX}{Example}

\DeclareMathOperator*{\argmax}{argmax}

\title{Lying Aversion and Vague Communication: \\An Experimental Study}
\author{
  Keh-Kuan Sun\footnote{Department of Economics, Fairfield University.  Email: ksun@fairfield.edu}
  \and
  Stella Papadokonstantaki\footnote{Department of Economics, Washington University in St. Louis.  Email: p.stella@wustl.edu}
  \thanks{We thank Guangying Chen for her contribution to the early version of this paper. We also thank Brian Rogers, Marta Serra-Garcia, Agne Kajackaite, Cynthia Cryder, John Nachbar, Jonathan Weinstein, Marcus Berliant, Sangmok Lee, Paulo Natenzon, John Rehbeck, Erik Kimbrough, Duk Gyoo Kim, Martin Dufwenberg, and seminar participants at Washington University in St. Louis, the ESA Job-Market Candidates Seminar Series, the Berlin School of Economics, the Korea Information Society Development Institute, the Tokyo Institute of Technology, Hitotsubashi University, the ECONtribute Summer Workshop 2021 on Social Image and Moral Behavior, the 2021 North American Summer Meeting of the Econometric Society, the 2021 Asian Summer Meeting of the Econometric Society, and the 2021 ESA Global Online Around-the-Clock Meetings. We would like to extend our sincere gratitude to the journal editor, the associate editor who handled our manuscript, and two anonymous referees for their invaluable feedback and constructive suggestions, which significantly improved the quality of this paper. We are grateful for the financial support from the Washington University in St. Louis and the Smith Institute for Political Economy and Philosophy of Chapman University.}
}

\date{\today}

\begin{document}
   	\maketitle\small
   	
   	\begin{abstract}
   	
   	An agent may strategically employ a vague message to mislead an audience's belief about the state of the world, but this may cause the agent to feel guilt or negatively impact how the audience perceives the agent. Using a novel experimental design that allows participants to be vague while at the same time isolating the internal cost of lying from the social identity cost of appearing dishonest, we explore the extent to which these two types of lying costs affect communication. We find that participants exploit vagueness to be consistent with the truth, while at the same time leveraging the imprecision to their own benefit. More participants use vague messages in treatments where concern with social identity is relevant. In addition, we find that social identity concerns substantially affect the length and patterns of vague messages used across the treatments.

    
   	\end{abstract}

    \vfill
    \pagebreak
\newgeometry{top=1.5in,bottom=1.5in,right=1.25in,left=1.25in}

\section{Introduction}

Communication has been studied extensively in economics in recent decades. In particular, the types of messages sent by an agent in the communication process and their impact on an audience's beliefs have been important subjects of game theory. Standard economic models predict that a sender will choose a message that yields the greatest benefit from the receiver's action, even when such a message is a lie. However, real-life observations consistently show that individuals often opt for evasiveness or vagueness instead of outright lies, even when lying could lead to greater personal benefits. For instance, politicians often employ evasive language or vague statements to avoid making explicit false claims or being labeled as outright liars.  They strategically choose words or phrases that allow for multiple interpretations or provide general statements without specific details. Similarly, doctors, who possess extensive knowledge about a patient's condition, may choose to communicate it in a more vague manner, especially when the diagnosis is severe or complex. These real-world scenarios raise two important questions: why and when does a sender prefer vagueness over profit-maximizing lies? Addressing these questions is crucial for developing a comprehensive understanding of misleading behavior in various applications, including problems associated with public good provision \autocite{Serra-Garcia2011}, sender-receiver disclosure games \autocite{Hagenbach2018}, and persuasion games \autocite{Deversi2021}. 

To explore these questions, it is crucial to examine the behavioral aspects of lying and vague communication. First, recent developments in the literature on lying behavior \autocite{Meibauer38, Fischbacher2013, Abeler2019} have shown that most people exhibit a non-trivial degree of lying aversion. If individuals have an intrinsic preference for honesty, the internal cost or guilt\footnote{Note that we use the term `guilt' specifically to refer to the internalized social norm of honesty. This is different from the usage in \textcite{battigalli_dufwenberg_2007} where the authors use the term to incorporate a broader notion of `letting others down.'} associated with lying would deter them from using outright lies and instead encourage the use of vague yet relatively truthful messages. Second, a message not only influences the audience's belief about the state of the world but also their perception of the agent's honesty. When individuals value their social identity as honest individuals, this external concern will impact the type of message they choose to communicate. The nontrivial cost of lying implies that communication is no longer a cheap talk, and misleading messages may bear strategic significance as credible signals about the agent's behavioral types to the audience. Conversely, such strategic incentives can also influence behavior. Thus, a rational agent must balance the degree of truthfulness and vagueness in the message they communicate.

In this paper, we present a model of a cheating game in which an agent may report a vague (set-valued) message or a precise (single-valued) message to an audience after privately observing the state of the world. We classify a message as truthful when it includes the true state of the world, while any message that fails to convey the true state is considered a lie. The agent’s utility depends on the monetary payoff and their truth-telling preferences. The monetary payoff is determined solely by the reported message. We study two separable motivations for honesty: the internal motivation for being honest and an external concern related to social identity for being seen as honest. We isolate the two costs by employing an anonymous environment where the audience cannot identify an agent with a message. Based on the model, we hypothesize that people use vague messages to reduce their internal guilt and increase their monetary payoffs. The supposed reduction of the social identity concern in the anonymous environment should lead to more straightforward profit-maximizing message choices, while the prediction is more opaque in the non-anonymous environment because of the multiplicity of equilibria introduced in a richer message space.

To test hypotheses pertaining to vague communication with lying aversion, we compare treatments from an online experiment in which subjects face a variant of the \textcite{Fischbacher2013} type of reporting task (hereafter ``FFH"). In this experiment, subjects privately observe an integer drawn randomly from a uniform distribution ranging from 1 to 10. The subjects are asked to report the number to the experimenter, and their monetary payment increases with the reported number. The basic idea of the experiment is that the discrepancy between the maximum numbers they could have reported and the actual numbers reported should capture subjects’ aversion to lying or, by the same token, their preferences for truth-telling. We generalize the FFH model by allowing subjects to transmit set-valued messages to understand the effect of vagueness in communication. In our setup, we allow subjects to be vague by reporting multiple numbers. The experimenter then chooses one number randomly from the reported numbers and pays the subject accordingly.

We employ a within \textemdash and between \textemdash subject design to test the predictions and, in particular, to distinguish between the intrinsic cost of lying and the social identity cost of being perceived as dishonest. Within each experimental session, subjects are asked to participate in two reporting tasks, with the observation process being identical and independent for both tasks. However, the set of available messages differs between the two tasks. In the "restricted communication" task, participants are limited to using only single-valued messages. In contrast, in the "unrestricted communication" task, subjects are allowed to utilize both single-valued and set-valued messages. To investigate the impact of the two distinct lying costs, we conducted different types of experimental sessions. We varied these sessions based on two factors: the anonymity of the subjects' identity and the observability of the random draw. In the non-anonymous treatment, each subject's response is known to the experimenter. Conversely, in the anonymous treatment, the responses are recorded under screen names, ensuring that the experimenter cannot link a subject's identity to their specific response. As a result, the subject's social identity concern, defined by how the experimenter perceives the subject's honesty based on the message they provide, remains constant throughout the experiment. By holding the social identity concern constant in the anonymous treatment, any variations in the subject's choices across different available message spaces can be attributed to changes in the internal cost of lying rather than the external social identity cost. We can then compare the anonymous environment with a non-anonymous counterpart to identify the effect of the social image concern in communication. Lastly, to comprehensively understand the role of the internal cost within the anonymous environment, we introduce a variation in the observability of the true state. By directly observing the true state, we can calculate the lying frequency rather than relying solely on statistical inference. 

The experimental data suggest that the majority of subjects utilize vague messages, and when vague messages are permitted, the reported average numbers are higher on average. However, the manner in which subjects employ vague messages varies between the different treatments. First, in the anonymous treatment, we note a decrease in the frequency of lying behavior when vague messages are permitted, as opposed to when they are not allowed. Considering that the social identity concern remains constant in the anonymous treatment, the observed decrease in lying behavior when vague messages are allowed suggests that employing a vague yet truthful message reduces the internal cost of lying. Yet, at the same time, they tend to report higher average numbers. This indicates that subjects exploit vagueness not only to avoid telling lies but also to leverage the imprecision to their advantage. Second, in the non-anonymous treatment, subjects use vague messages more frequently compared to the anonymous counterpart.
The higher frequency of vague messages in the non-anonymous treatment, together with the relevance of social identity concerns, indicates that a message's vagueness influences message choices by impacting both the internal and external costs of lying. Furthermore, we find that the pattern of reported messages differs substantially between the two types of sessions. In the anonymous treatment, when subjects use vague messages, the majority do not avoid the most obvious forms. Specifically, they tend to report a combination of their true observations and the maximum number (10). However, in the non-anonymous treatment, a much smaller fraction of subjects use such obvious messages. This difference may arise from the fear that obvious vagueness might be interpreted as lies by the audience, resulting in a potential discredit of their social identities.

Our paper serves as a bridge between the literature on lying behavior and a broader array of studies that delve into vague communication. Existing literature on vague communication focuses on the strategic use of vague messages without considering the behavioral aspect of lying aversion or explicitly studying its nature. For instance, \textcite{Serra-Garcia2011}, analyzing a public-goods provision game between two agents with information asymmetry, assume that a set-valued vague message would incur a much lower lying cost than a precise, single-valued outright lie. Our experiment tests this assumption and provides further insight into the cost of lying with respect to the social identity concern as well. \textcite{agranov2012ignorance} show that, in coordination games with multiple equilibria, it might be beneficial for a benevolent sender to use vague communication. \textcite{zhang2022delegation} show that, in a delegation game, social welfare is higher when messages are intervals. \textcite{wood2022communication} concludes that transmitted information is more accurate when senders have the option to send either precise or vague messages. \textcite{Deversi2021} study the strategic use of vagueness in a voluntary disclosure game. They allow subjects to send an interval that contains their type, and this formal message structure is similar to that in our design. However, our design differs from theirs in two crucial ways. Firstly, our design offers subjects increased message choice flexibility, enabling them to select any subset of the state space. Additionally, we introduce a between-subject variation on anonymity. The integration of message choice flexibility and anonymity variation offers us the opportunity to investigate how attitudes towards lying and misleading behavior influence the use of vague messages in conjunction with strategic motives.

Conceptually, we view lying behavior as an optimization process in which a rational agent chooses the optimal degree of dishonesty, by balancing the marginal monetary benefit of lying with the non-monetary costs that depend on the likelihood that an individual is found to be lying. In a more specific context involving the nature of lying behavior, \textcite{Sobel2020} provides a comprehensive framework in which to understand lying and deception in games. They distinguish between three properties \textemdash the form, the interpretation, and the consequence \textemdash  of a message, and our model closely follows the framework suggested in that paper. Within this framework, we particularly acknowledge the notions of malleable lies discussed in \textcite{Turmunkh2019} and deniable lies in \textcite{Tergiman2022} as they bear a close relation to our empirical findings concerning the use of vague messages. However, in contrast to the definitions of malleable and deniable lies, which rely on interpretations of messages, we define lying strictly by reference to the formal property of a message: a message is a lie only when its sender knows that it does not map to the true state of the world. The agents in our model face both monetary and, potentially, psychological consequences for their message choices. Through an analysis of equilibrium strategies, we provide a basis to understand how the interpretation of a message affects an agent's decision. This structure allows us to decompose the internal cost of lying from the concern for social identity.

Structurally, we study lying behavior within the framework of signaling games. Since \textcite{Crawford1982}, a stream of research has tested the assumption that lying is costless, which implies that individuals will lie whenever there is a material incentive to do so. However, empirical evidence continues to challenge this notion, as studies such as those by \textcite{Dickhaut1995}, \textcite{Blume1998}, \textcite{Gneezy2005}, \textcite{Sanchez-Pages2007}, \textcite{shalvi2011justified}, and many more, consistently demonstrate that individuals exhibit aversion to lying and accept significantly lower payoffs than what theories have predicted.

An empirical consensus regarding the FFH experiments is that people do not lie to the extent that they could, preferring to tell only minor lies, potentially because lying is costly. As the monetary payoff in such an experiment is independent of the drawn number, subjects should report non-maximal numbers only if aversion to lying is present, for otherwise this becomes a simple case of a cheap-talk game. \textcite{Abeler2019} combine data from 90 such experiments that describe the average reporting behavior. They find that the behavior is indeed bounded away from the maximal report but also departs from a complete truth-telling scenario. In addition, by ruling out other explanations for truth-telling which are popular in the literature (e.g. inequality aversion or seeking a reputation for not being greedy), they conclude that a preference for being seen as honest and a preference for being honest are the main motivations. As \textcite{Abeler2019} note in their conclusion, however, the FFH paradigm has focused on subjects reporting a single number and excludes lies by omission or vagueness. Our paper extends the FFH paradigm by allowing vague communication using set-valued messages and contributes to our knowledge of how a message space in communication plays a role in an agent's reporting decision when lying entails a cost.

The studies that are closest to our work are \textcite{Gneezy2018} and \textcite{Khalmetski2019}. In these studies, an agent cares not only about obtaining a monetary payoff but also about whether lying takes place as well as how others interpret their report, a finding that is consistent with the empirical findings of the aforementioned study by \textcite{Abeler2019}. These papers adopt the FFH paradigm and simplify the sender/receiver structure of a signaling game into a cheating game between reporting agents and an observing audience. This idea, in conjunction with the assumption of a non-atomic game, minimizes the role of the receiver and allows a sharper focus on that of the sender. Moreover, this simplification makes it possible to develop a unique characterization of off-path beliefs in the context of mild conditions.  While we inherit their assumptions of these assumed motivations for truth-telling \textemdash internal cost and the external concern with one's social identity \textemdash extending the message space to include set-valued messages yields more opaque predictions with multiple equilibria and off-path beliefs. We overcome this difficulty by introducing an anonymous environment to hold the impact of social identity constant.

The remainder of the paper is organized as follows. In Section 2 we define our terms and present the model setting. Section 3 provides theoretical analysis, while in Section 4 we present the experiment hypotheses based on the theoretical predictions. In Section 5 we describe our experiment design and procedure, in Section 6 we summarize the experimental outcomes, and Section 7 concludes. We list all the proofs and additional details regarding the experiment in the appendices.

\newpage

\section{Model}
	\subsection{A model of lying aversion with vague communication}
	We study lying aversion with vague communication by considering a variant of the \textcite{Fischbacher2013} cheating game with a population of agents and one audience. An agent privately observes the state of the world $i \in \Omega$, where $\Omega = \{1,2, \dotsc, N\}$ is finite. We assume $i$ is drawn i.i.d. from a uniform distribution over $\Omega$ across agents. Each agent has a private type $t$ that represents their intrinsic aversion to lying. We also assume $t$ is i.i.d. across agents. Its CDF $F(t)$ is strictly increasing, continuous, and has support $[0,T]$. Denote an agent who observes the state $i$ and has the intrinsic aversion type $t$ as a type $(i,t)$ agent.

	An agent reports a message $J$ to the audience after observing the true state of the world $i$. This reporting takes place only once and there are no repeated interactions. Like \textcite{Gneezy2018}(GKS) and \textcite{Khalmetski2019}(KS), we assume that an agent's utility consists of three components: a monetary payoff, an internal concern with being honest, and a social identity for being seen as honest. We use the term `social identity' to distinguish it from the typical understanding of reputation in a dynamic game, as we are modeling a one-shot game. Formally, we write a type $(i,t)$ agent's utility for reporting a message $J$ as:
	\begin{equation}
		U(i,J,t) = \underbrace{\bar{\pi}(J)}_{\text{monetary payoff}} - \underbrace{\mathds{1}(i \not \in J) \left[ t + c(\pi(\{i\}), \bar{\pi} (J)) \right]}_{\text{internal guilt}} + \underbrace{\gamma \rho(J)}_{\text{external social identity}},
		\label{eq:utility}
	\end{equation}
    and we define each component of the utility function like the following.

	The message $J$ is a nonempty\footnote{An empty set is often interpreted as silence and plays an interesting role in the literature on vague communication. We abstract away from silent messages to compare cases where lying aversion is accompanied by vague language with cases where lying aversion is accompanied by precise language.} subset of the state space: $J \in M^\Omega \equiv 2^{\Omega} \setminus \emptyset$. An important distinction between our model and GKS and KS is that in our model the message space admits set-value messages instead of being isomorphic with the state space.  This generalization allows messages to be categorized in multiple ways. We define the relevant terms below.

	\begin{DF}
	    A message $J$ is truthful if $i \in J$. A message is a lie if it is not truthful.
	\end{DF}

	\begin{DF}
	    A message is called precise if it is a singleton set; otherwise it is vague.
	\end{DF}

    \begin{sloppypar}
    Let $M^{\Omega}_{P} \equiv \{ \{1 \}, \{2 \}, \dotsc, \{N\}\}$ and $M^{\Omega}_{V} \equiv M^\Omega \setminus M^{\Omega}_{P}$ denote the sets of precise messages and vague messages, respectively. For example, if $\Omega=\{1, 2, 3\}$, the set of possible precise messages is $\{\{1\}, \{2\}, \{3\}\}$, and the set of possible vague messages is $\{ \{1, 2\}, \{2, 3\}, \{1, 3\}, \{1, 2, 3\} \}$.\footnote{The fully vague message $\Omega \in M^\Omega$ is analogous to the notion of evasive lying in \textcite{Khalmetski2017jpube} by which the sender falsely states to have not observed the state of the world.} Note that our definition of lying depends purely on the form of a message, as in the framework proposed in \textcite{Sobel2020}. The consequences of a message choice are reflected through effects on the components of the agent's utility.
    \end{sloppypar}

	The monetary payoff for an agent maps a message to the agent's utility: $\pi:M^\Omega \rightarrow \mathbb{R}$. For simplicity, we assume that $\pi(J)$ is a uniform draw over $J$ and that the agent is a risk-neutral, expected-utility maximizer. Let us denote $\bar{\pi} (J) = E[\pi(J)]$. Note that, when a message is precise, i.e. $J = \{x \}$ with $x \in \Omega$, the monetary payoff $\pi(J)$ is simply $x$.

	In addition to receiving a  monetary payoff, the agent also has two distinct motivations for being honest. First, the agent has an internal motivation for being honest. When the agent's  report $J$ is not truthful\footnote{In our model, we make the assumption that the cost of lying is independent of the precision of a message, which may seem strong at first glance. However, note that our primary goal in this study is not to quantify the internal cost of lying, but rather to confirm its existence. To this end, the discrete assumption is adequate for the purpose.}, that is, $i \not \in J$, the dishonesty incurs an internal cost,  $t + c(\pi(\{i\}), \bar{\pi} (J))$. The agent's private type $t$ captures their sensitivity to the intrinsic (fixed) cost of lying. The function $c(\pi(\{i\}), \bar{\pi} (J)): \mathbb{R} \times \mathbb{R} \rightarrow \mathbb{R}$ represents the variable cost of lying depending on the size of the lie. The size of the lie is measured as the ex-ante difference between the monetary payoff for a report $J$ and that of the true and precise report $\{i\}$. We assume that i) $c(\cdot) \geq 0$;  ii) $c(\pi(\{i\}),\pi(\{i\}))=0$; iii) $c(\pi(\{i\}), \bar{\pi} (J))$ is weakly increasing in $|\pi(\{i\}) - \bar{\pi} (J)|$; iv) $c(\pi(\{i\}), \pi(\{i\})+1)<1$;\footnote{This assumption excludes the trivial case where the variable cost of lying is so high that no agent chooses to lie.} and v) $c(\pi(\{i\}),\bar{\pi} (J))+c(\bar{\pi} (J),\bar{\pi} (K)) \geq c(\pi(\{i\}),\bar{\pi} (K))$.

    The agent also has an external motivation to be ascribed a  social identity for being seen as honest; that is, the agent's utility depends on the audience's belief about how honest the agent is. We assume that the audience is a rational Bayesian who forms a posterior belief based on the agent's report $J$.  This belief, in turn, depends on the agent's mixed strategy in equilibrium.

    \begin{DF}
    A type $(i,t)$ agent's mixed strategy is a mapping $\sigma: \Omega \times [0,T] \rightarrow [0,1]^{2^{N} - 1}$ such that
		\begin{equation}
			\sigma(i, t) = \left( \sigma^{\{1\}}_{it}, \sigma^{\{2\}}_{it}, \dotsc , \sigma^{\{1,2, \dotsc, N\}}_{it} \right)
			\label{eq:mixed_strategy}
		\end{equation}
	where $\sigma^{J}_{it}$ is the probability which the intrinsic aversion type $t$ agent with true observation $i$ assigns to the report $J \in M_\Omega$.
    \end{DF}

    In equilibrium, the audience's posterior belief about whether an agent's report $J$ is truthful is computed using Bayes' rule:
		\begin{equation}
			\rho(J) = \frac{P(\text{agent is honest} \land \text{agent reports } J)}{P(\text{agent reports } J)} = \frac{\sum_{k \in J}(\int^{T}_{0} \sigma^{J}_{kt} d f(t) )}{\sum^{N}_{k=0}(\int^{T}_{0} \sigma^{J}_{kt} d f(t) )}
			\label{eq:posterior}
		\end{equation}

	We normalize the posterior belief $\rho$ in terms of the agent's utility by parameter $\gamma$. The parameter measures the agent's sensitivity to their social identity reflected in reporting message $J$. We further assume that i) $\gamma$ is homogeneous across agents and is common knowledge; moreover, ii) $N + \gamma < T$. The latter condition ensures that $F(N+\gamma)<1$, or there always exists a positive mass of agents who will be truth-telling for any observation $i$.

	Because the agent's payoff depends on the audience's belief, we adopt the notion of sequential equilibria in an induced psychological game in the sense of \textcite{Dufwenberg2018}, \textcite{Battigalli2009} and \textcite{Geanakoplos1989}.

	An equilibrium is a set of mixed strategies and beliefs satisfying the following conditions:
	\begin{equation}
	    \forall(i, J, t): \sigma^{J}_{it} >0 \text{ only if } J \in \argmax_{J^{\prime}} U(i, J^{\prime}, t),
	\end{equation}
	\begin{equation}
	    \forall(i, t): \sum_{J \in M^{\Omega}} \sigma^{J}_{it} = 1,
	\end{equation}
	\begin{equation}
	    \forall J: \rho(J) = \frac{\sum_{k \in J}(\int^{T}_{0} \sigma^{J}_{tk} d f(t) )}{\sum^{N}_{k=0}(\int^{T}_{0} \sigma^{J}_{tk} d f(t) )}.
	\end{equation}

	As noted by GKS, the existence of an equilibrium follows \textcite{Schmeidler1973} in treating each type $(i, t)$ as a player.

	\subsection{The communication environment and the anonymity of agents}
	The baseline model assumes no restriction on message space $M^\Omega$. If we restrict message space to $M^{\Omega}_{P}$, on the other hand, we obtain a model with restricted communication. We refer to the baseline model as the model with unrestricted communication. At the risk of abusing the notation, let us denote a message as $j$ in the model with restricted communication.

	Furthermore, for a more comprehensive understanding of the relationship between the use of vague messages and the costs of lying, it is helpful to isolate their effects on internal guilt from their effects on external social identity. We can achieve this by employing an anonymous environment where the audience cannot identify an agent with a message. This implies that the agent's report does not alter the audience's belief, and hence the agent's social identity, now labeled as $\rho_0$, remains constant independent of their reporting choice in the anonymity-preserving  environment. On the other hand, we call an environment non-anonymous when the audience can associate a message with its sender. Formally, we write an agent's utility in the anonymous environment as:
	\begin{equation}
		U_{A}(i,J, t) =\bar{\pi} (J) - \mathds{1}(i \not \in J) \left[ t + c(\pi(\{i\}), \bar{\pi} (J))\right] + \rho_0.
		\label{eq:utility_anonymous}
	\end{equation}

	We thus create four distinct environments by varying the restriction on the message space and the anonymity of the agents. In these environments, the agent is non-anonymous with restricted (precise only) communication (NA-R), non-anonymous with unrestricted (potentially vague) communication (NA-UR), anonymous with restricted communication (A-R), and anonymous with unrestricted communication (A-UR). Note that the NA-R environment corresponds to the standard FFH model where an agent can send only precise messages and the audience can identify a message's sender.
	
\begin{table}[]
\centering
\resizebox{0.5\textwidth}{!}{%
\begin{tabular}{ccc}
\hline
                         & Restricted & Unrestricted \\ \hline
\vspace*{-10pt} &  &  \\
Non-Anonymous  &NA-R & NA-UR \\
Anonymous & A-R  & A-UR  \\ \hline
\end{tabular}%
}
\caption{Four environments}
\label{tab:treatments}
\end{table}

    \section{Analysis}
    In this section we describe the equilibria and agents’ behavior in each of the four environments as defined above. We begin by arguing that we should expect to observe vague messages used at positive probabilities in any equilibrium.
    
    \begin{theoremEnd}[proof at the end, text link= ]{proposition}
				In any equilibrium under unrestricted communication, there exist types of agents who use at least one vague message with positive probability in their mixed strategies.
				\label{pr:use_vague}
		\end{theoremEnd}
		    \begin{proof}
		          See Appendix A.
		    \end{proof}
			\begin{proofEnd}
			    \hfill
			    \begin{enumerate}[{Case }1.]
			        \item Anonymous environment: as shown in Proposition \ref{pr:A-UR_report}, all truth-tellers use the optimal vague message.
			        \item Non-anonymous environment: suppose there exists an equilibrium where no agent uses a vague message. Let $\rho(j)=0$ for all $j \in M^{\Omega}_{V}$. Given that all messages used with positive probability are precise, we know all precise messages are used with positive probability in this equilibrium. We use Lemma \ref{lm:NA-R_lie} and \ref{lm:NA-R_no-under} to argue that there exists a positive probability that agents lie upward. Let $l^{*}$ be the threshold defined in Lemma \ref{lm:NA-R_threshold}.

				Let us first argue that there exists agents who observe $l^{*}-1$ and lie by reporting $l^{*}$. Suppose not. Then there must exist some $l > l^{*}$ such that
				\begin{align*}
					U(l^{*}-1, l, t)-U(l^{*}-1, l^{*}, t) = (l- l^{*}) + (c(l^{*}-1, l^{*}) - c(l^{*}-1, l)) + \gamma(\rho(l)- \rho(l^{*})) >0.
				\end{align*}
				Note that $c(l^{*}-1, l^{*}) - c(l^{*}-1, l) \leq 0$ because $l > l^{*}$ and $c$ is increasing in the distance between the two arguments. Also, $c(l^{*}-1, l^{*}) - c(l^{*}-1, l) \leq c(i, l^{*}) - c(i, l) \leq 0$ for all $i < l^{*}$ because of the triangular inequality assumption. This implies that for all agents whose true observation is below $l^{*}$ is better off by reporting $l$ instead of $l^{*}$. As no agent would lie by reporting $l^{*}$, this is a contradiction to the definition of threshold.

				Now consider the agent who observes $l^{*}-1$ and lies by reporting $l^{*}$. The agent receives the utility of
				$$
					U(l^{*}-1, l^{*}, t) = l^{*} - (t + c(l^{*}-1, l^{*})) + \gamma \rho (l^{*}).
				$$
				However, if the agent reports $\{l^{*}-1, l^{*}+1\}$ instead, the agent receives
				$$
					U(l^{*}-1, \{l^{*}-1, l^{*}+1\}, t) = l^{*}.
				$$
				That is, the agent is better off by reporting $\{l^{*}-1, l^{*}+1\}$ when
				$$
					t  > \gamma \rho (l^{*}) -c(l^{*}-1, l^{*}),
				$$
				which happens with a positive probability. Furthermore, the above analysis is valid for all $\rho(\{l^{*}-1, l^{*}+1\}) \geq 0$. This is a contradiction to the assumption of an equilibrium with no vague messages. Therefore, there exists a positive probability that agents use a vague message in any equilibrium in the NA-UR environment.
			    \end{enumerate}
				
			\end{proofEnd}
        
        The intuition behind Proposition \ref{pr:use_vague} is that there exist agents with moderately high intrinsic lying aversion type $t$ who would prefer to be truthful while increasing the monetary payoff. Therefore, in their messages, they include higher numbers apart from their true observations. For these agents, such messages strictly dominate any precise lie.
        
        In comparison with the unrestricted communication environment in which we are  interested, a particularly useful element when analyzing a restricted communication environment is the observation that there always exists a positive mass of truth-tellers for each observation $i$ due to the assumption about the distribution of the intrinsic aversion $t$. While this assumption is sufficiently plausible in many contexts, GKS and KS showed that it aids us especially in uniquely characterizing the audience's posterior belief in all equilibria.\footnote{For more details, please refer to the proof of Lemma 1 in \textcite{Gneezy2018} where the posterior belief $\rho$ pertaining to all equilibria must be unique when the social identity matters, or $\gamma >0$ in our model specification.} Thus, we are able to make sharp equilibrium predictions even in the case of the non-anonymous environment. This, however, is unfortunately not the case when we relax the message space to allow vague messages: the richer message space induces multiplicity in agents’ reporting strategies and the audience's off-path beliefs. We illustrate this point with the following examples.

		\begin{EX}
            \label{ex:full_vagueness}
				Consider the case where $\gamma>\frac{N-1}{2}$. Then this case in which all agents report the vaguest messages $\{1, 2, \dotsc, N\}$, such that $\rho(J) =1$ only when $J = \{1, 2, \dotsc, N\}$ and $0$ otherwise, is an equilibrium.
		\end{EX}

		This example describes the scenario in which agents care greatly about social identity and are forced to stick to the (exogeneously) given norm of full vagueness. While this specific example is similar to situations that might occur in the real world,\footnote{For instance, we may imagine a group of politicians all replying with the same vague message to a politically sensitive question because they know that any message other than a fully vague message  will be interpreted as a lie and hurt their social identity. Many similar examples can arise in situations where agents value their social identity highly. We appreciate Brian Rogers for suggesting this interpretation of the equilibrium.} an important implication is that there can exist a multiplicity of equilibria depending on the combination of $\gamma$ and the off-path beliefs. Consider the following example.

		\begin{EX}
            \label{ex:equilibrium_interval}
				All agents report a message in the form of an interval: $[i, N] = \{i, i+1, \dotsc, N-1, N\}$, and the audience assigns a positive posterior belief only to those interval-messages and zero to all other messages. We need the following conditions to hold to constitute such equilibria:
				\begin{itemize}
					\item truth-tellers: $\frac{i+N}{2} + \gamma \rho(J) > U(i, \{ i, x^{*}, x^{*} +1, \dotsc, N \} ,t) $\footnote{This message maximizes the expected payoff conditional on including the true state $i$. We later define this kind of message as an optimally vague message.} for all $i$ and an interval message $J$
					\item liars: $\frac{k+N}{2} - \mathds{1}(i \not \in J)(t + \bar{\pi}(J)) + \gamma \rho(J) > U(i, J^{\prime}, t)$ for any interval message $J$ and non-interval message $J^{\prime}$,
				\end{itemize}
                where $k$ is the smallest number of the interval message $J$.
				The conditions hold trivially when $\gamma \rightarrow \infty$, and the signaling game becomes isomorphic with the NA-R scenario, where each message $i$ is simply replaced with the interval $[i, N]$. The ending sequence conveys no information to the audience, and the equilibrium follows that of the NA-R.
		\end{EX}
        
        The multiplicity of equilibria obscures our understanding of the agents’ behavior regarding vague communication. We thus first turn our attention to anonymous environments to delineate the relationship between one’s internal cost of lying and vague messages.
        
        Because we are now free of the social identity component, we can utilize the fact there is a clear mapping between an agent’s type $(i, t)$ and the resulting behavior. That is, fixing the observation $i$, one’s lying behavior simplifies to a monotone function of $t$ in the A-R environment: the agent reports $j > i$ when $( j - i) + c(i, j) > t$. Thus, there exists a threshold $t^{*}_{i}$ for each observation $i$ below which the agent reports a lie and above which the agent tells the truth. This threshold, however, becomes much higher in the A-UR environment because now the agent is equipped with vague messages that allow them to remain truthful while increasing the expected payoff. This fact leads to the following observations.
        
        \begin{OBS}
                If a type $(i,t)$ agent reports truthfully after observing state $i$ in the A-R environment, then the agent also reports truthfully when observing $i$ in the A-UR environment.
                \label{lm:A-R_true-A-UR_true}
        \end{OBS}
        
        \begin{OBS}
                For each observation $i<N$, there exists a positive mass of agents who lie in the A-R environment but report truthfully in the A-UR environment.
                \label{lm:A-R_lie-A-UR_true}
        \end{OBS}

        The immediate corollary of these two observations is the following proposition, which states that the mass of liars is greater when communication is restricted in the anonymous environments.

		\begin{theoremEnd}[proof at the end, text link= ]{proposition}[A-R/A-UR]
		    The set of types (i,t) of agents who lie in equilibrium in the A-UR environment is a subset of liar types in any equilibrium in the A-R environment. The expected monetary earnings are greater on average when the communication is not restricted.
                \label{pr:A-R-A-UR}
		\end{theoremEnd}

        \begin{sloppypar}
        Another implication of the anonymous environment worth noting is that the absence of the social identity concern simplifies the problem into a straightforward comparison between the solutions to truth-telling-constrained and unconstrained optimizations. That is, for each type $(i,t)$ agent, there exists both $\max_{J: i \in J} U(i, J, t)$ and $\max_{J \in M^{\Omega}} U(i, J, t)$, and the agent reports truthfully only if the two maximums coincide.    
        \end{sloppypar}

        We can add structure to the constrained optimization. Intuitively, this maximum has the form of a union of the true observation $i$ and some ending sequence $x, x+1, \dotsc, N$. We can solve the optimization problem to find the threshold $x^{*} = \lceil (N+2) - \sqrt{2N-2i+3} \rceil$. \footnote{We can simplify the optimization problem by approximating it with a continuous uniform distribution: $\argmax_{x} \int^{i+1}_{i}\frac{u}{N-x+2}du + \int^{N}_{x}\frac{u}{N-x+2}du$}
        
        \begin{DF}
           The optimal vague message (OVM) for each true observation $i$ is defined as
            \begin{equation}
                OVM_i \equiv \{ i, x^{*}, x^{*} +1, \dotsc, N \},
                \label{def:ovm}
            \end{equation}
            where $x^{*}$ the threshold above which including the values maximizes the expected monetary payoff conditional on using a truthful message.
        \end{DF}
        
        Then $OVM_i \in \argmax_{J: i \in J} U(i, J, t)$ and weakly dominates all truthful messages. The weak dominance is because of the discrete nature of the underlying uniform distribution and risk neutrality. For example, an agent who observes $i=8$ is indifferent between reporting $\{8, 10\}$ and $\{8, 9, 10\}$. This allows us to make the following inferences in the A-UR environment:

            \begin{theoremEnd}[proof at the end, text link= ]{proposition}
				In the A-UR environment, 
                    \begin{enumerate}[i.]
                    \item all truth-tellers earn $\bar{\pi} (OVM_i)$;
					\item no agent's message contains a number below the true observation.
					\item no precise message except $\{N\}$ is truthful.
				\end{enumerate}
    				\label{pr:A-UR_report}
            \end{theoremEnd}
            \begin{proof}
               See Appendix A.
            \end{proof}
            \begin{proofEnd}
                \begin{enumerate}
                    \item By the definition of the OVM, it is the message that maximizes the utility of an agent who observes state $i$ conditional on including the true state $i$ in their message. If $\argmax_{J: i \in J} U(i, J, t)$ is a singleton set, we are done. If not, the agent may use another message in $\argmax_{J: i \in J} U(i, J, t)$, but the expected earning is still the same.
                    \item Suppose an agent reports a message $J$ with $\min J < i$. Then the message $J \cup \{i\} \setminus \{ \min J \}$ strictly dominates $J$.
                    \item Suppose an agent truthfully reports $\{i\}$. If $i = N$, we are done. If not, the message is strictly dominated by $OVM_i$.
                \end{enumerate}
            \end{proofEnd}
        
        Lastly, we conclude this section by comparing the behavior of an agent in the non-anonymous environment with that of an agent in the anonymous environment. A thought experiment that involves  choosing an agent and comparing their behavior in the two environments easily leads to the conjecture that the absence of the social identity concern should only encourage more lies. The following lemma shows that this is indeed the case under restricted communication.
        
        \begin{theoremEnd}[proof at the end, text link= ]{lemma}
                If a type $(i,t)$ agent lies in an equilibrium in the NA-R environment, then the agent lies in the A-R environment.
                \label{lm:NA-R_then-A-R}
            \end{theoremEnd}
            \begin{proof}
                See Appendix A.
            \end{proof}
            \begin{proofEnd}
                Let $j>i$ be a part of an equilibrium strategy for an $(i,t)$-agent in the NA-R environment:
                \begin{equation*}
                    U(i,j,t)=j-[t+c(i,j)]+\gamma \rho(j) \geq U(i, j^{\prime}, t) \quad \forall j^{\prime} \in M^{\Omega}_P.
                \end{equation*}

                We can infer that
                \begin{equation*}
                    j-[t+c(i,j)]+\gamma \rho(j) \geq i + \gamma \rho(i),
                \end{equation*}
                or
                \begin{equation*}
                    (j-i)  - [t + c(i,j)] \geq \gamma(\rho(i) - \rho(j)).
                \end{equation*}

                As no agent underreports in the A-R environment, it suffices to show that this agent does not tells the truth in the A-R. Suppose not.
                \begin{equation*}
                    U_A (i, i, t) = i > U_A (i, j, t) = j - [t + c(i,j)] .
                \end{equation*}

                This implies $\rho(i) < \rho(j)$, meaning the agent lied in the NA-R both because there were a monetary gain and a social identity gain.
                \begin{align*}
                    (j-i) + [c(i^{\prime},i) - c(i^{\prime},j)] + \gamma(\rho(j) - \rho(i)) & \geq  c(i^{\prime},i) - c(i^{\prime},j) + t + c(i,j) >0
                \end{align*}
                because of the triangular inequality assumption: $c(i, i^{\prime}) + c(i^{\prime}, j) \geq c(i,j)$.
                As the choice of this agent is arbitrary, this is the case for all agents who do not observe $i$ never lies by reporting $i$; in turn, this implies $\rho(i)=1$, a contradiction. Therefore, the agent must lie in the A-R if the agent sees the same observation $i$.
            \end{proofEnd}

            Using the above lemma, we now generalize the argument to compare the probability that an arbitrarily chosen agent lies in the two environments.

            \begin{theoremEnd}[proof at the end, text link=]{proposition}[NA-R/A-R]
                The set of types $(i,t)$ of agents who lie in an equilibrium in the NA-R environment is a subset of the set of liar types in any equilibrium in the A-R environment. The expected monetary earnings are higher  on average in the A-R environment.
                \label{pr:NA-R-A-R}
            \end{theoremEnd}
            \begin{proof}
               See Appendix A.
            \end{proof}
            \begin{proofEnd}
                WTS $P(\text{lie}_{NA-R}) < P(\text{lie}_{A-R})$.

				We argue by the law of total probability:
				\begin{align*}
					P(\text{lie}_{NA-R}) & = \sum^{N}_{i=1} P(\text{observe $i$}) P(\text{lie}_{NA-R} | \text{observe $i$}); \\
					P(\text{lie}_{A-R}) & = \sum^{N}_{i=1} P(\text{observe $i$}) P(\text{lie}_{A-R} | \text{observe $i$}).
				\end{align*}
					Because the probability of observing some $i$ is uniform in both environments, it suffices to show that the conditional probability of lying in A-R is greater than or equal to that in NA-R for all true observation $i$.

				We learned that there exists some threshold $1<l^{*}<N$ in Lemma \ref{lm:NA-R_threshold}. Thus, conditional on that an agent observing $i \geq l^{*}$, we have the conditional probability of the agent reporting a lie as
				$$
					P(\text{lie}_{NA-R} | \text{observe $i$}) = 0;
				$$
				for all intrinsic aversion type $t$, while
				$$
					P(\text{lie}_{A-R} | \text{observe $i$}) >0
				$$
				because of the positive probability that the agent has the type $t$ small enough to report a lie.

				Now consider the case of $i < l^{*}$. Let $T_i$ be a subset of $\mathbb{R}$ such that
				\begin{align*}
					P(\text{lie}_{NA-R} | \text{observe $i$}, t \in T_i) > 0; \\
					P(\text{lie}_{NA-R} | \text{observe $i$}, t \not \in T_i) = 0.
				\end{align*}

				By Lemma \ref{lm:NA-R_then-A-R}, $P(\text{lie}_{A-R} | \text{observe $i$}, t \in T_i) = 1$; and $P(\text{lie}_{A-R} | \text{observe $i$}, t \not \in T_i) \geq 0$. Thus, regardless of $P(t \in T_i)$, we have
				$$
					P(\text{lie}_{NA-R} | \text{observe $i$}) \leq P(\text{lie}_{A-R} | \text{observe $i$})
				$$
				for all $i < l^{*}$. Therefore, $P(\text{lie}_{NA-R}) < P(\text{lie}_{A-R})$.

                Also, because the monetary payment is a monotone mapping of the reports under the restricted communication, and because any lying takes the form of reporting upward, agents would earn more monetary payoff on average as the probability of lying is greater in the A-R.
            \end{proofEnd}

\section{Hypotheses}

In this section, we have transformed our theoretical analysis into specific hypotheses that can be tested in an experiment. The results of the experiment, which will be discussed in the following section, will be based on these hypotheses.
        
   	\begin{HP}
	    In both the NA-UR and A-UR environments, agents use vague messages.
	    \label{hp:use_vague}
	\end{HP}
	Hypothesis \ref{hp:use_vague} is a consequence of Proposition \ref{pr:use_vague}.

	\begin{HP}[A-R/A-UR]
        In the anonymous environment,
        \begin{enumerate}[i.]
            \item more agents lie when communication is restricted (precise): $\text{lie}_{A-R} \geq \text{lie}_{A-UR}$;
            \item an agent who is truthful in A-R is also truthful in A-UR conditional on the same observation;
            \item some agents who lie in A-R report truthfully in A-UR conditional on the same observation;
            \item agents earn higher monetary payoffs on average when communication is not restricted (vague): $\text{earning}_{A-R} \leq \text{earning}_{A-UR}$.
        \end{enumerate}
        \label{hp:A-R-A-UR}
	\end{HP}
	Hypothesis \ref{hp:A-R-A-UR} is a consequence of Proposition \ref{pr:A-R-A-UR} and the above observations about behavior in the A-UR environment.

	\begin{HP}
	    In the A-UR environment,
	   \begin{enumerate}[i.]
			\item all truth-tellers in A-UR use OVM;
			\item no message contains a number below the true observation;
			\item no precise message except $\{N\}$ is truthful.
		\end{enumerate}
		\label{hp:A-UR_report}
	\end{HP}
	Hypothesis \ref{hp:A-UR_report} is a consequence of Proposition \ref{pr:A-UR_report}.

	\begin{HP}[NA-R/A-R]
        Under restricted communication, agents earn higher monetary payoffs on average in the anonymous environment: $\text{earning}_{NA-R} \leq \text{earning}_{A-R}$.
        \label{hp:NA-R-A-R}
	\end{HP}
	Hypothesis \ref{hp:NA-R-A-R} is a consequence of Proposition \ref{pr:NA-R-A-R}.

\section{Experimental Design and Procedure}
\subsection{Experimental Design}
Our experimental treatments vary along three dimensions: 1) we consider precise or vague messages, 2) we vary the experimenter's ability to identify responses from an individual subject (anonymity), and 3) we vary the experimenter's ability to know the true observation (observability). We introduce variation in subjects' anonymity and observability of the true observation through different types of experiments. Within each session, a subject confronts two stages of reporting tasks that represent the availability of vague messages. In each stage, subjects are incentivized to observe a random number and later report the number to the experimenter. The subjects' earnings depend on the number or numbers they report.

Our experiment encompasses a total of six treatments, achieved through a combination of between-subject and within-subject variations. These treatments are labeled as NA-R\footnote{The NA-R treatment is equivalent to the original experiment conducted by \textcite{Fischbacher2013}.}, NA-UR, AO-R, AO-UR, A-R, and A-UR. Table \ref{tab:experimental_treatments} presents an overview of our experimental treatments.

\begin{table}[htbp]
\centering
\resizebox{0.7\textwidth}{!}{%
\begin{tabular}{ccc}
\hline
                         & Restricted & Unrestricted \\ \hline
 \vspace*{-10pt} &  &  \\
Non-Anonymous \& Unobservable & NA-R & NA-UR \\
Anonymous \& Unobservable & A-R  & A-UR  \\
Anonymous \& Observable   & AO-R & AO-UR \\ \hline
\end{tabular}%
}
\caption{Experimental Treatments}
\label{tab:experimental_treatments}
\end{table}

In the anonymous sessions, responses are recorded under screen names so that the experimenter cannot map a subject’s identity to their response. As the subjects are instructed that the experiment prohibits such mapping by design, this treatment should establish the effect of suppressing the social identity concern and emulate the environment where $\gamma \rightarrow 0$. In the non-anonymous (identifiable) session, on the other hand, the experimenter knows each subject’s response.

The ‘stage’ is our basic unit of observation. During each stage, subjects are presented with a random integer that has been generated uniformly between 1 and 10. They are then asked to report the number they observed.  In the treatment where the true observation is observable, the random integer is generated within the experimental software, allowing the experimenter to have knowledge of the true observation. Conversely, in the treatments where the true observation is not observable, the subjects use an external website, which we provided as a link to a Google search result for the phrase "random number between 1 and 10," to generate the random integer. This ensures that the experimenter is unable to access the true observation.

The observation process for both tasks within each session is identical, however, the set of messages that can be utilized by the subjects differs between the tasks. In the restricted stage, only single-valued messages are permitted, while in the unrestricted stage, both single-valued and set-valued messages are allowed. Subjects can send such messages by clicking numbered boxes on their screens. Please refer to Figures \ref{fig:precise_screenshot} and \ref{fig:vague_screenshot} in appendix C for screenshots of the experiment software. In the restricted stage, subjects receive payment in dollars equivalent to half of the reported number. In the unrestricted stage, subjects are paid in dollars equivalent to half the value of the reported number if they report a single number. If they provide multiple numbers, the computer will randomly choose one, and the payment will be equal to half the value of the selected number in dollars. To emulate the one-shot game structure of our model, we asked subjects to participate only once in each of the stages and we randomized the order of precise and vague stages within a session.\footnote{Using OLS regressions, we find that there is no order effect on the average reports.}

\subsection{Experimental Procedure}
We recruited 231 student subjects from the subject pool of the Missouri Social Science Experimental Laboratory (MISSEL) at Washington University in St. Louis. An experiment session is composed of instructions, a preliminary quiz, and two choice task stages. Subjects are invited through the MISSEL's ORSEE system \autocite{Greiner2015} and, once registered, they receive a link to a Zoom meeting. Upon joining the Zoom room, the experimenter reads the instructions and conducts a preliminary quiz to ensure subjects comprehend the procedures. Subjects are allowed to participate in the main experiment only after successfully passing the quiz, with a maximum of three attempts. The experimenter then shares the main experiment website link via Zoom's chat window. The main experiment was implemented using the Qualtrics online survey platform.

In anonymous sessions, both the experimenter and subjects keep their videos off, and attendance is not explicitly recorded. To maintain anonymity, subjects use a screen name, which the experimenter clarifies is for data analysis purposes only and cannot be linked to their identity. In non-anonymous sessions, the experimenter and subjects have their videos on, and attendance is documented using the experiment roster.

For anonymous sessions, the experimenter bulk-purchases Amazon gift cards corresponding to each screen name's earnings. They share a list of gift codes and associated screen names with a third party (a staff member at Washington University in St. Louis), deliberately omitting the monetary values connected to each code. This process ensures anonymity, as the experimenter can map between responses and screen names but not between screen names and true identities. The third party can only map between true identities and screen names, not between true identities and their corresponding responses. In non-anonymous sessions, subjects submit their email address and taxpayer information to receive the Amazon gift card directly from the experimenter.

We conducted 24 sessions and each session lasted for approximately 30 minutes, including the instructions read during Zoom meetings, a screening quiz to make sure that subjects understood  the experimental procedures, and two stages of the main experiment task via Qualtrics. \footnote{Initially, we conducted two types of sessions: Non-Anonymous sessions where the true observation was not observable and Anonymous sessions where the true observation was observable. The first 11 sessions were held between August and October 2020, and seven sessions were conducted in September 2022. The results of a logit regression analysis revealed no significant differences in the probability of lying between these two groups of sessions. In April 2023, we conducted additional Anonymous sessions where the true observation was not observable. } In all cases, subjects received a \$2 show-up fee, so the total amount they could earn ranged from \$2 to \$12, with an average total fee, including the show-up fee, of \$9.96. No subject participated in more than one session. See Appendix C for our experimental instructions.

\section{Results}

    We present basic summary statistics in Table \ref{tab:basic_result}. The study included 10 non-anonymous sessions where the true observation was not observable, 8 anonymous sessions where the true observation was observable, and 6 anonymous sessions where the true observation was not observable. On average, the non-anonymous sessions had 8.2 participants, anonymous sessions with observable observations had 12.3 participants, and the anonymous sessions with unobservable observations had 9.1 participants.\footnote{Participants chose time blocks voluntarily, and we provided the anonymous version of the software in larger sessions.}. In the unrestricted communication treatments, we used the mean of the numbers included in each subject's report when calculating the average report. We computed the average length of vague messages using the average number of numbers included in the messages.

   In the non-anonymous treatment, the average reported numbers were 7 and 8.282 for the restricted and unrestricted treatments, respectively. In the anonymous treatment with observable observations, the average reported numbers were 7.619 and 8.132, while in the anonymous treatment with unobservable observations, the average reported numbers were 8.163 and 8.490, respectively. These reported numbers are significantly lower than those in the profit-maximizing reports and are in line with findings reported in the literature that have highlighted agents' preference for truth-telling. Overall, we find that allowing vague messages increased the mean of numbers reported. \footnote{We also find that the probability of lying is independent of gender in both the restricted and unrestricted treatments. Additional details regarding the logit regression outcomes are available in Appendix B.} Furthermore, we find a statistically significant difference between the non-anonymous restricted (NA-R) and anonymous restricted (A-R) treatments (p-value = 0.017), while the differences between the other treatments were not statistically significant. These results suggest that anonymity has a significant impact on lying behavior when subjects are only allowed to communicate using precise messages. Finally, we show that the introduction of non-anonymity significantly affects the choice to send vague messages. In the treatment where social identity concern is relevant, subjects use vague and longer messages more often.

\begin{figure}[H]
  \begin{subfigure}{0.45\textwidth}
    \centering
    \includegraphics[width=\linewidth, height=6cm]{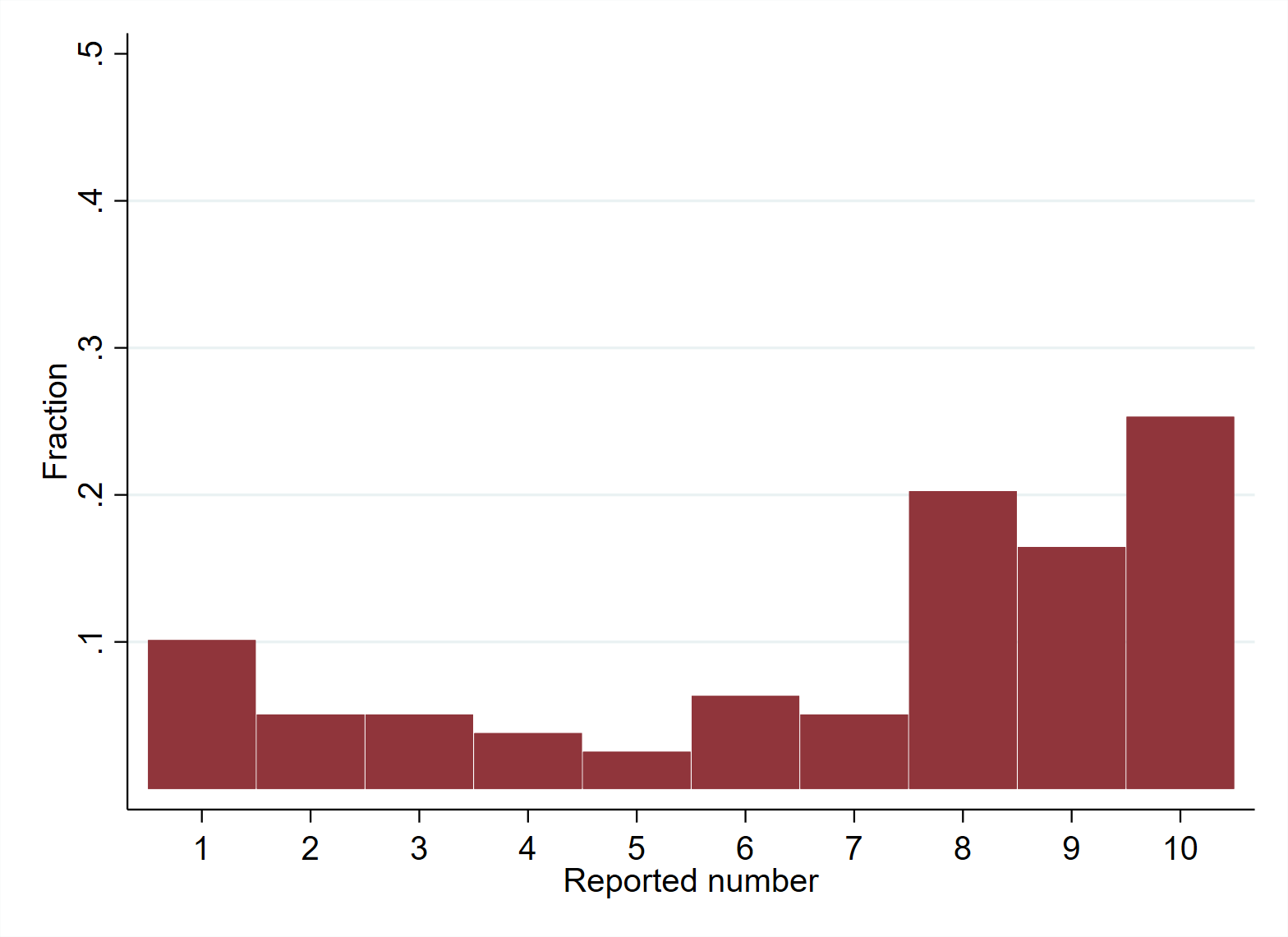}
    \label{fig:NA_R}
    \caption{NA-R}
  \end{subfigure}
  \hfill
  \begin{subfigure}{0.45\textwidth}
    \centering
    \includegraphics[width=\linewidth, height=6cm]{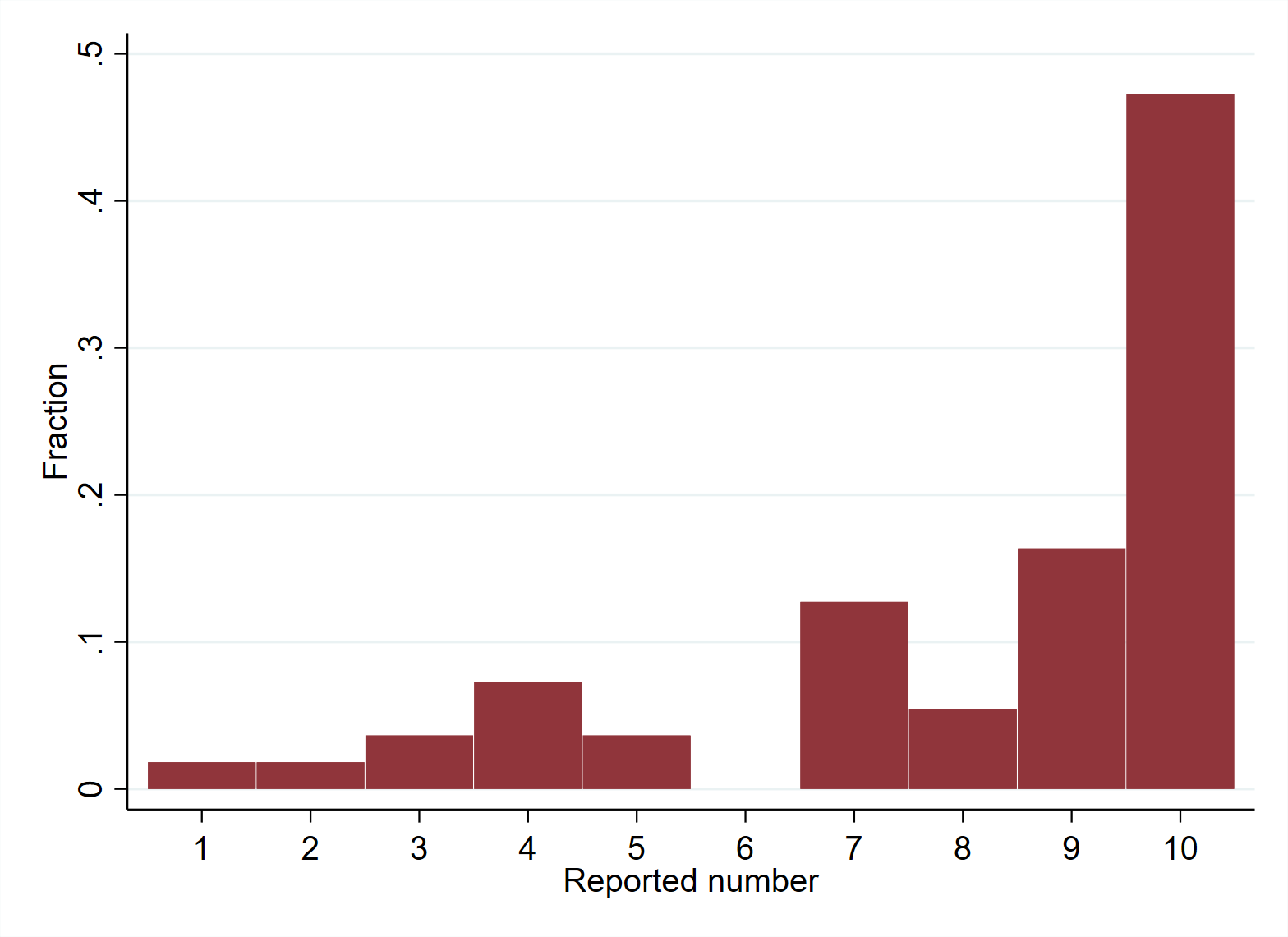}
    \label{fig:A_R}
    \caption{A-R}
  \end{subfigure}
  \hfill
  \begin{subfigure}{0.45\textwidth}
    \centering
    \includegraphics[width=\linewidth, height=6cm]{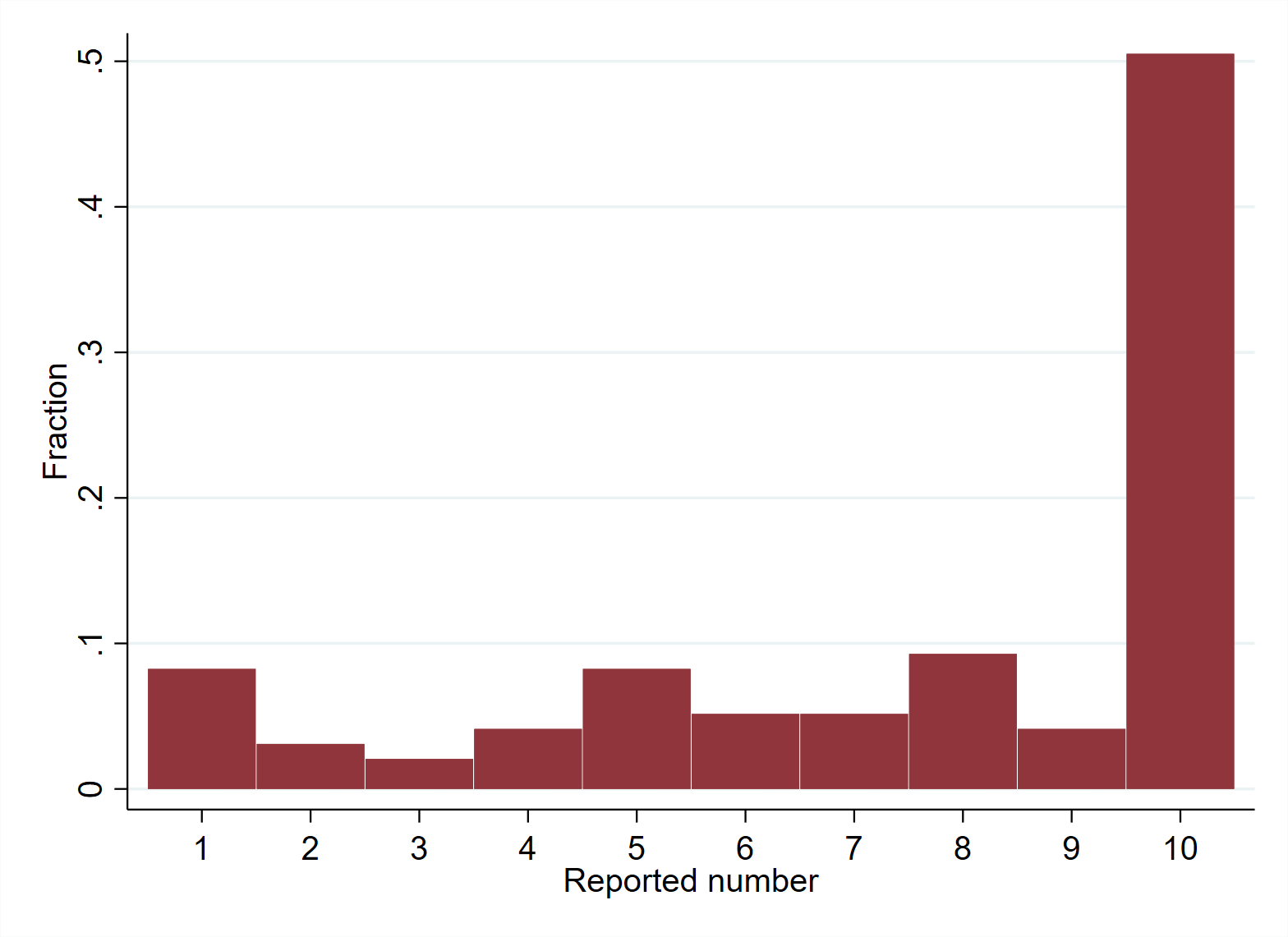}
    \label{fig:AO_R}
    \caption{AO-UR}
  \end{subfigure}
  \caption{Fractions of reported numbers in NA-R, A-R, and AO-R}
  \label{fig:hist_restr}
\end{figure}

\begin{figure}[H]
  \begin{subfigure}{0.45\textwidth}
    \centering
    \includegraphics[width=\linewidth, height=6cm]{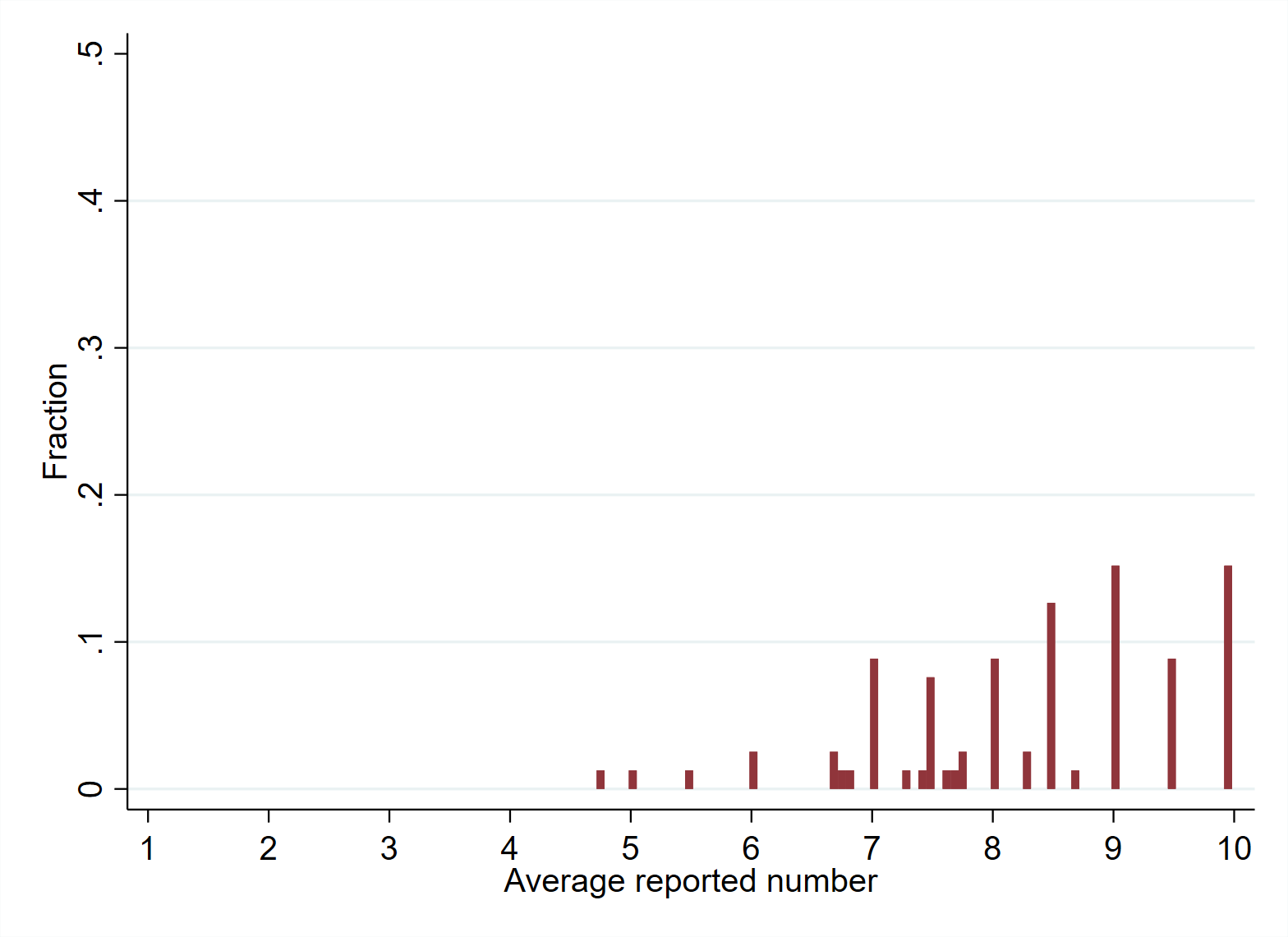}
    \label{fig:NA_UR}
    \caption{NA-UR}
  \end{subfigure}
  \hfill
  \begin{subfigure}{0.45\textwidth}
    \centering
    \includegraphics[width=\linewidth, height=6cm]{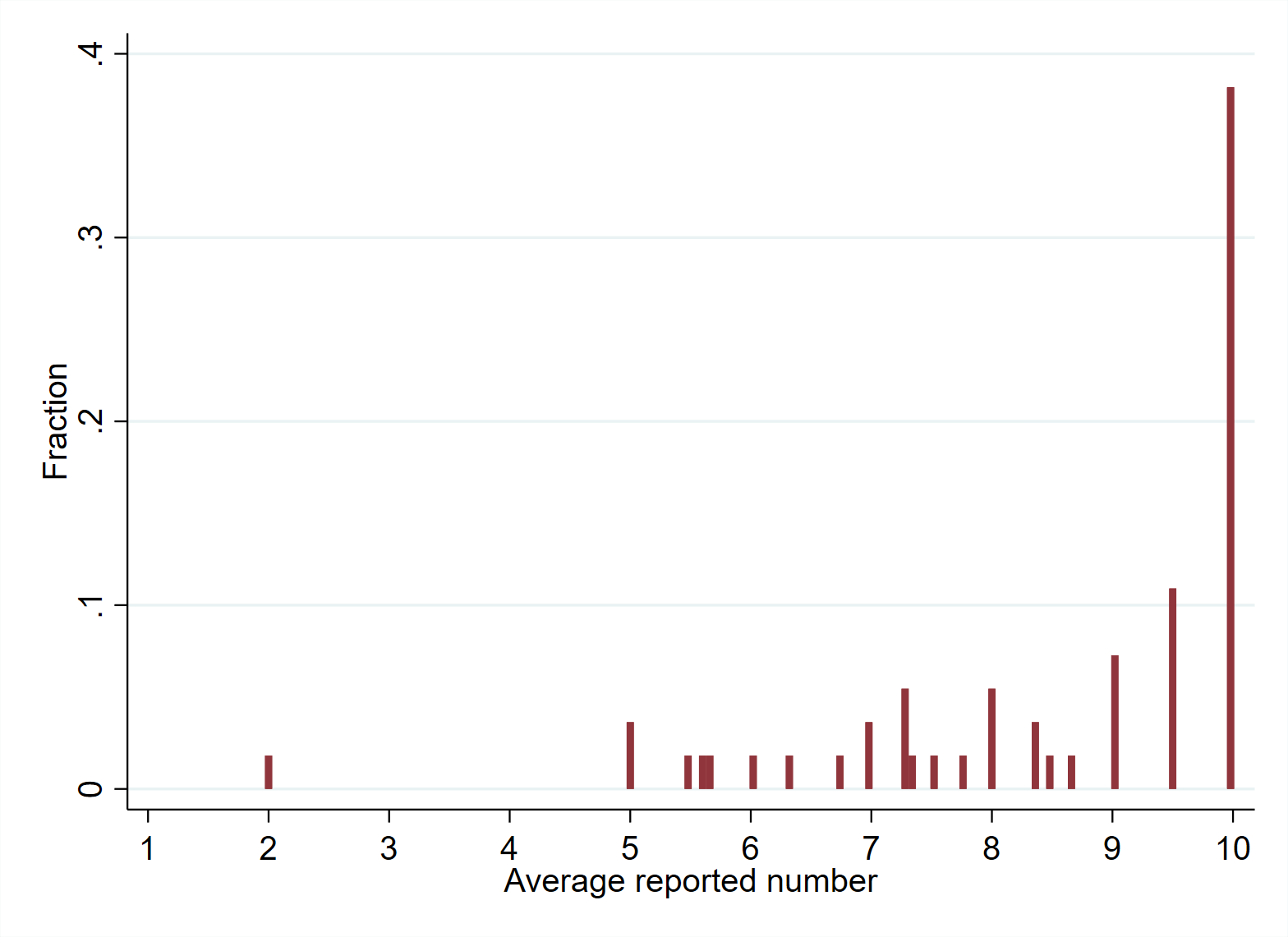}
    \label{fig:A_UR}
      \caption{A-UR}
  \end{subfigure}
  \hfill
  \begin{subfigure}{0.45\textwidth}
    \centering
    \includegraphics[width=\linewidth, height=6cm]{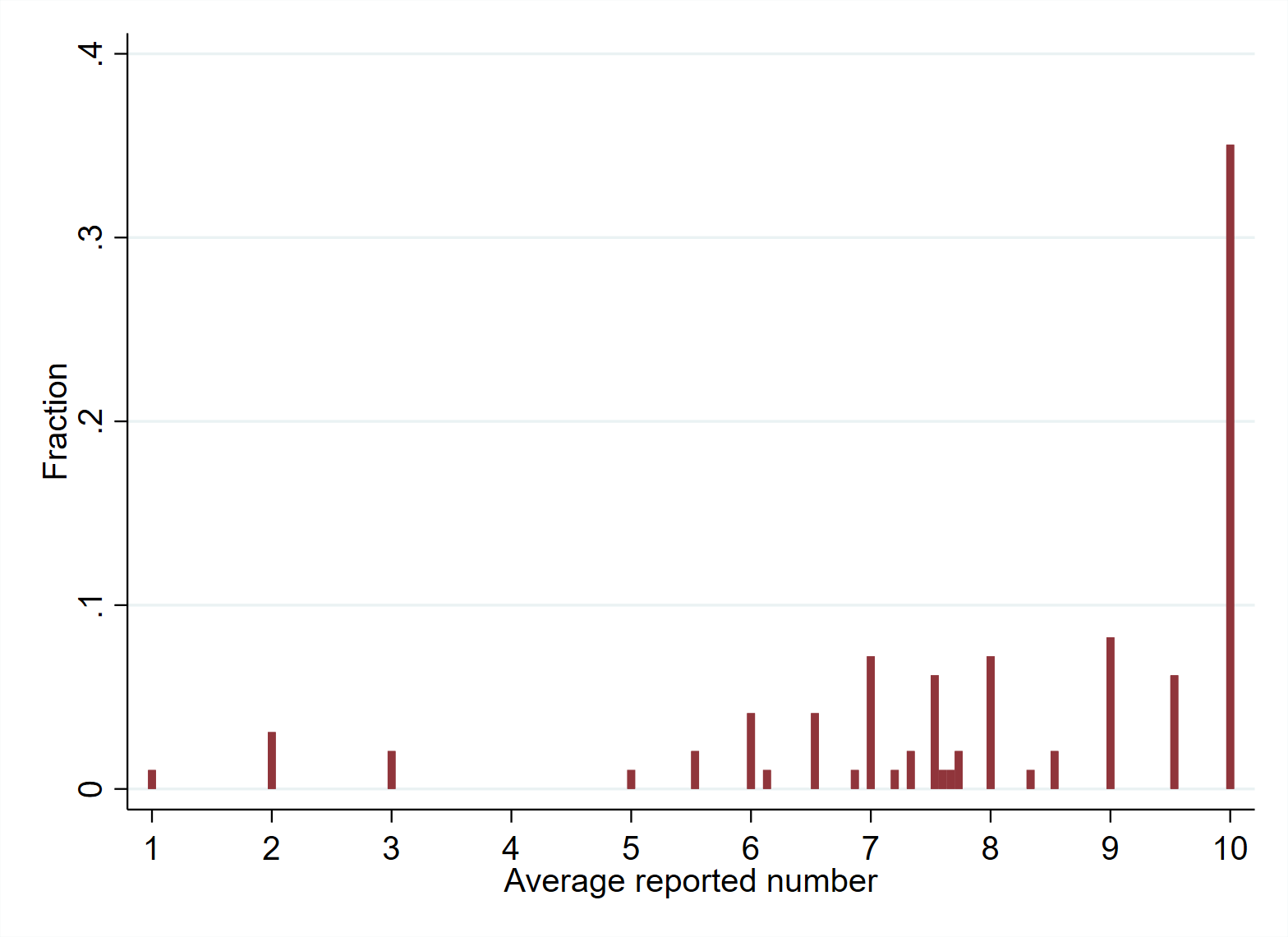}
    \label{fig:AO_UR}
      \caption{AO-UR}
  \end{subfigure}
  \caption{Fractions of average reported numbers in NA-UR, A-UR, and AO-UR}
  \label{fig:hist_unrestr}
\end{figure}

    Figures \ref{fig:hist_restr} and \ref{fig:hist_unrestr} depict histograms illustrating the frequency of reported messages for each treatment. Figure \ref{fig:hist_restr} that represents the result from NA-R treatment is analogous to that of the original FFH paper. By analyzing the histograms of the Non-Anonymous treatment in comparison to the Anonymous treatments under restricted communication, it becomes evident that a greater proportion of subjects opted for the values 8 and 9, rather than 10, as opposed to the Anonymous treatments. This observation aligns with previous studies in the literature, which have indicated that individuals are more likely to engage in partial deception when social identity concerns are relevant. 

\begin{table}[hbt!]
    \centering
 
\resizebox{0.95\textwidth}{!}{%
    \begin{tabular}{lcccccc}
    \toprule
    \multicolumn{1}{c}{} & \multicolumn{1}{c}{Restricted} & \multicolumn{3}{c}{Unrestricted} & \multicolumn{1}{c}{Observations} \\ \midrule
                         & Average                        & Average                          & Vague (\%)                    & Length                        &                               \\ \cmidrule(lr){2-5}
    Non-anonymous \& Unobservable & 7.000                        & 8.282                            & 75\%                          & 2.810                         & 79                            \\
    Anonymous \& Observable   & 7.619                        & 8.132                            & 48\%                          & 2.103                         & 97                            \\ 
    Anonymous \& Unobservable  & 8.163                        & 8.490                            & 51\%                          & 2.054                         & 55                            \\ 
    
    \bottomrule
    \multicolumn{1}{l}{Difference in means test} & & & & & \\
     \midrule
     Test (NA = AO) &   0.619         &   0.149       &           &   0.707***  \\
      & (0.465)& (0.256)& & (0.237)\\
     Test (NA = A)  &     1.164**      &    0.209      &           & 0.756***   \\
     & (0.483)& (0.280)& &(0.249)\\
     Test (AO = A)  &    0.545       &   0.358       &         & 0.049\\
     & (0.457)& (0.323)& &(0.245)\\
      \bottomrule
    \end{tabular}
    }
    \caption{Data summary}
    \label{tab:basic_result}
\end{table}

In the following paragraphs we present the primary findings of our study.
    
    \begin{RS}
        In the NA-UR environment, 59 of 79 participants (74.7\%) used vague messages. In the AO-UR environment, 47 of 97 participants (48.4\%) used vague messages, and 28 of 55 participants (50.9\%) in the A-UR.
    \end{RS}
Our findings indicate that a substantial proportion of subjects chose to utilize vague messages when given the choice. This observation provides support for Hypothesis \ref{hp:use_vague}, which proposes that individuals would employ vague messages in both Anonymous and Non-anonymous settings. The main idea is that certain agents, characterized by a moderately high aversion to lying, prefer to be honest while also maximizing their monetary gains.  To achieve this, they use vague messages containing higher numbers along with their true observations.

    \begin{RS}
       In the AO environment,
        \begin{enumerate}
            \item more participants lied when communication was restricted (43.3\% in AO-R and 30.9\% in AO-UR);
            \item almost all participants who were truthful in AO-R remained truthful in AO-UR;\footnote{Four subjects were truthful in the AO-R and lied in the AO-UR, but their true observations were much larger in the AO-UR than in the AO-R treatments. It is likely that the counterfactual would be consistent with our prediction conditional on observing similar numbers in both treatments.};
            \item 52.58 \% of subjects reported truthfully under both AO-R and AO-UR, 26.80\% always lied, while 16\% switched from lying to truth-telling when they were allowed to be vague;
            \item participants reported higher numbers on average when communication was not restricted (7.619 in AO-R and 8.282 in AO-UR).
        \end{enumerate}
    \end{RS}

\begin{figure}[H]
  \begin{subfigure}{0.45\textwidth}
    \centering
  \includegraphics[width=\linewidth, height=6cm]{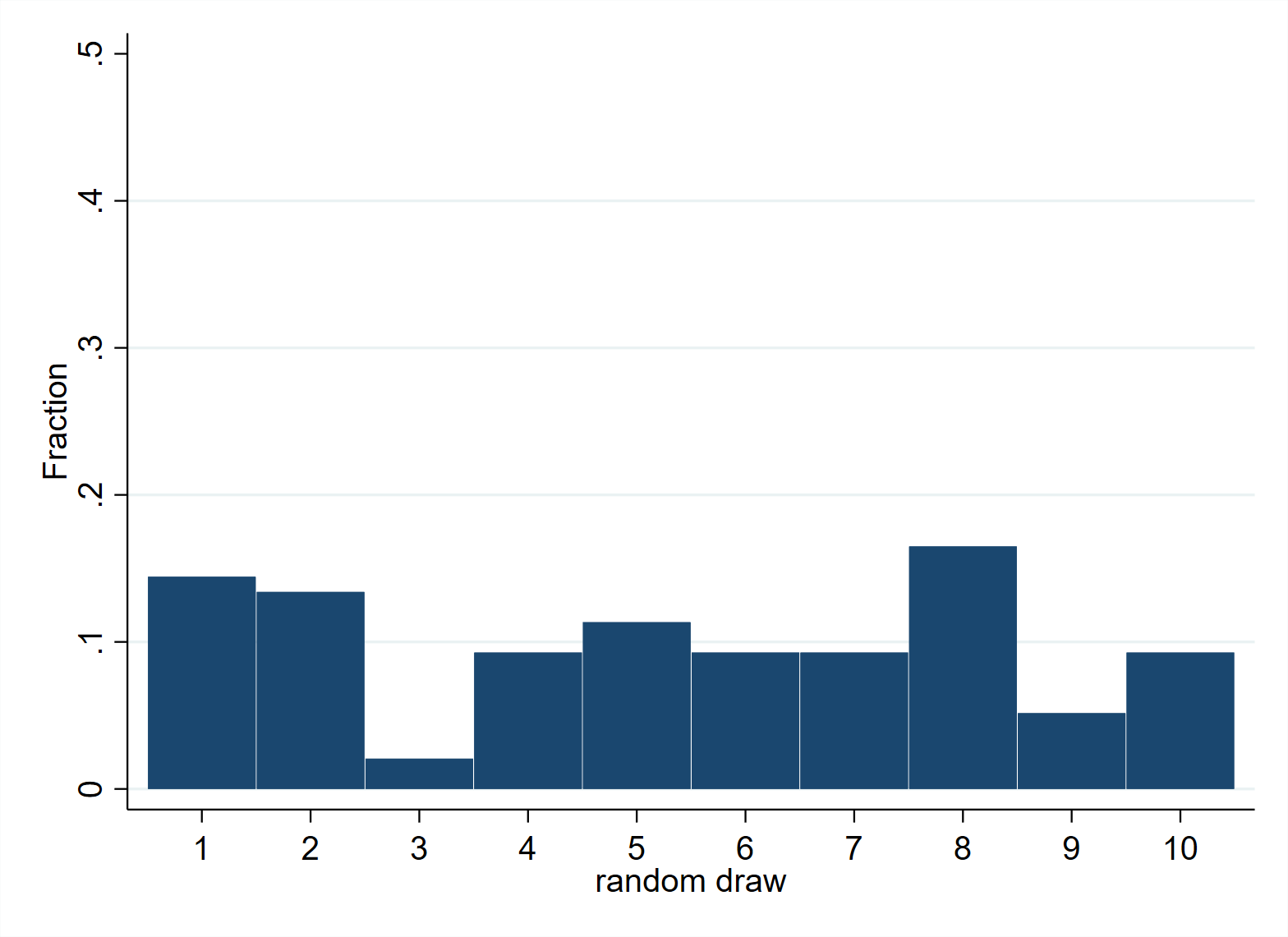}
    \phantomsubcaption %
    \label{fig:NA_R}
  \end{subfigure}%
  \hfill
  \begin{subfigure}{0.45\textwidth}
    \centering
     \includegraphics[width=\linewidth, height=6cm]{figures/hist_AO_R.png}
    \phantomsubcaption %
    \label{fig:A_R}
  \end{subfigure}
  \caption{Fractions of random draws and reported numbers in AO-R}
  \label{fig:draw_restr}
\end{figure}

\begin{figure}[H]
  \begin{subfigure}{0.45\textwidth}
    \centering
       \includegraphics[width=\linewidth, height=6cm]{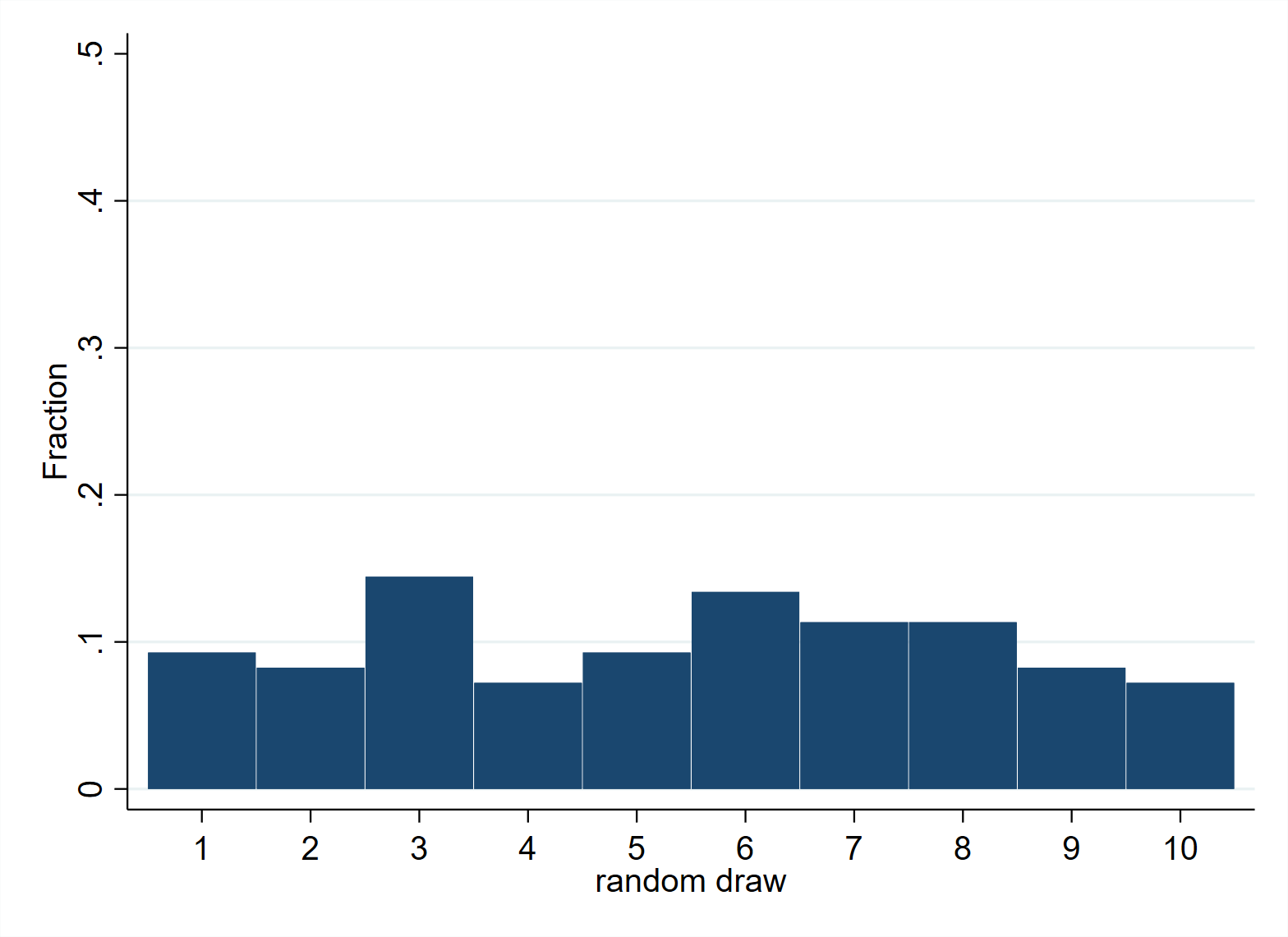}
    \phantomsubcaption %
    \label{fig:NA_R}
  \end{subfigure}%
  \hfill
  \begin{subfigure}{0.45\textwidth}
    \centering
    \includegraphics[width=\linewidth, height=6cm ]{figures/hist_AO_UR.png}
    \phantomsubcaption %
    \label{fig:A_R}
  \end{subfigure}
  \caption{Fractions of random draws and reported numbers in AO-UR}
  \label{fig:draw_unrestrict}
\end{figure}

Figures \ref{fig:draw_restr} and \ref{fig:draw_unrestrict} present histograms illustrating the distribution of random draws and actual reports within the AO-R and AO-UR environments, respectively. A comparison between the results obtained with restricted and  unrestricted communication in anonymous sessions confirms Hypothesis \ref{hp:A-R-A-UR}. More subjects lied in the restricted than in the unrestricted treatment and their behavior can be classified into three types: truth-tellers (reported truthfully under both treatments), conditional liars (switched from lying to truth-telling when they were allowed to be vague) and liars (lied under both treatments). This is in line with previous work on lying aversion (\textcite{Gneezy2018}, \textcite{Khalmetski2019}, among others). An interesting point is that 15 of the 16 participants who switched from lying to truth-telling reported picking the number 10 in AO-R, yet they chose vague and truthful messages when the expected earnings were lower than 10.  This provides evidence that using vague messages decreases the internal cost of lying.

Table \ref{tab:Anonymous_Treat} presents the average numbers reported in the AO-R and AO-UR treatments for truth-tellers, conditional liars, and liars. We find that truth-tellers report on average 1.282 higher numbers when they are allowed to be vague. This result is significant at the 1\% significance level and is consistent with the model's prediction that truth-tellers seek payoff maximization conditional on including true observations in their reports. Conditional liars reported on average 1.156 lower numbers when they were allowed to be vague. This result is significant at the 1\% significance level and supports Hypothesis \ref{hp:A-R-A-UR}. We interpret this result as implying that subjects with moderate $t$ prefer to use vague messages and include true states to reduce the internal cost of lying. Consistent with our expectation regarding agents with higher $t$, we did not find a significant impact of unrestricted communication on the behavior of liars, as indicated by their average reports. In addition, t-tests of the difference between subjects' true observations and their reports show over-reporting across all three types in the AO-UR treatment. The main intuition behind this is that agents with sufficiently low lying aversion type (liars) always prefer saying maximal lies, while agents with sufficient high lying aversion type (truth-tellers and conditional liars) exploit vagueness to be consistent with the truth while at the same time they try to increase their monetary gains by incorporating higher numbers into their messages.

\begin{table}[hbt!]
\centering
\scalebox{1}{
\resizebox{0.9\textwidth}{!}{%
\begin{tabular}{@{}ccccc@{}}
\toprule
{\color[HTML]{333333} } &
  {\color[HTML]{333333} Restricted} &
  {\color[HTML]{333333} Unrestricted} &
  \makecell{\color[HTML]{333333} Differences in means} &
  {\color[HTML]{333333} Observations} \\ \midrule
{\color[HTML]{333333} Truth-tellers} &
  {\color[HTML]{333333} 5.608} &
  {\color[HTML]{333333} 6.890} &
  {\color[HTML]{333333} 1.282***} &
  {\color[HTML]{333333} 51} \\
  &   &  &    (0.464)       &    \\

\begin{tabular}[c]{@{}c@{}}Conditional liars\end{tabular} & 9.875     & 8.718    & -1.156*** & 16 \\
 &   &  &    (0.370)       &    \\

{\color[HTML]{333333} Liars} &
  {\color[HTML]{333333} 10} &
  {\color[HTML]{333333} 9.962} &
  {\color[HTML]{333333} -0.038} &
  {\color[HTML]{333333} 26} \\
       &   &  &    (0.027)       &    \\                  
  \midrule
   
Test (Report = True)                                                   &           &          &           &    \\ \midrule
Truth-tellers                                                          & 0         & 1.953*** &           & 51 \\
                                                                       &           & (0.267)  &           &    \\
\begin{tabular}[c]{@{}c@{}}Conditional liars\end{tabular} & 5.000*** & 1.531*** &           & 16 \\
                                                                       & (0.791)   & (0.432)  &           &    \\
Liars                                                                  & 5.462***  & 5.076*** &           & 26 \\
 
                                                                       & (0.521)   & (0.543)  &           &    \\ 
                                                  
  \bottomrule
\end{tabular}%
}
}
    \caption{Anonymous treatment with observable random draw}
    \label{tab:Anonymous_Treat}
\end{table}

Furthermore, our analysis reveals a significant negative correlation between the true observation and the probability of lying in the restricted communication environment. In contrast, no significant relationship is observed in the unrestricted communication environment.  Table \ref{tab:logit} presents the estimated coefficients of the logit model of the effect of true observations on the probability of lying. This result is intuitive, as a subject who randomly drew a low observation can increase their payoff only by lying when the communication is restricted to precise messages. These subjects can, however, always employ a vague yet truthful message in the unrestricted environment; hence the true observation has no effect.

\begin{table}[hbt!]
\centering
\resizebox{13cm}{!}{%
\begin{tabular}{ccc}
\hline
{\color[HTML]{333333} }         & {\color[HTML]{333333} Observation (Restricted)} & {\color[HTML]{333333} Observation (Unrestricted)} \\ \hline
\\
\multicolumn{1}{l}{Lying dummy (p-value)} & 0.055                                         & 0.321                                            \\
                                          \\ \hline
\end{tabular}%
}
\caption{Logit model of the probability of lying}
    \label{tab:logit}
\end{table}

    \begin{subtheorem}{RS}
        \begin{RS}
            In the AO-UR environment,
            \begin{enumerate}
                \item
                    \begin{enumerate}
                        \item 28.4\% (19 of 67) of truth-tellers reported the optimal messages (either the OVM or the honest 10), 26.9\% (18 of 67) reported a pair of true observations and 10;
                        \item 23.9\% (16 of 67) of truth-tellers used a precise message below $\{10\}$;
                    \end{enumerate}
                \item only 2 of 97 subjects included a number below the true observation in the report;
                \item all precise messages reporting numbers below $\{10\}$ were truthful.
            \end{enumerate}
        \end{RS}
    \end{subtheorem}

    \begin{figure}[ht]
        \centering
        \includegraphics[width=0.95\textwidth]{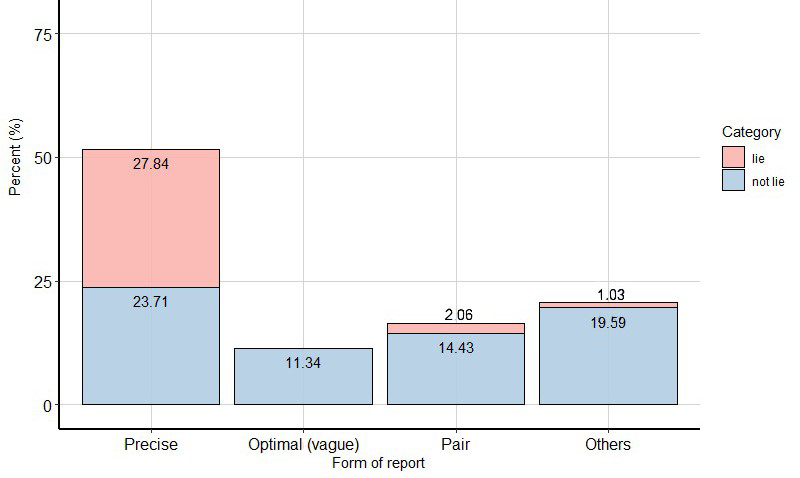}
        \caption{Message types used in AO-UR treatments}
        \label{fig:report_form_AO-UR}
    \end{figure}

    This result partially supports Hypothesis \ref{hp:A-UR_report}. Figure \ref{fig:report_form_AO-UR} summarizes the types of messages used in AO
    -UR treatments. Overall, 51.5\% (50 of 97) of participants used precise messages, while 48.5\% (47 of 97) used vague messages. Among the 50 precise messages, 54\% (27 of 50) were lies and 46\% (23 of 50) were not lies. All the liars reported drawing the maximum of 10. Among the 23 precise truth-tellers, 7  observed 10 and reported so.
    
    The model predicts that all truth-tellers seek payoff maximization conditional on including true observations in their reports. If we combine both optimal vague messages (including honest 10 as the optimal message) and the pair-type messages into a broader set of payoff-increasing truthful messages, we find that the majority (55.2\%) of truth-tellers maximized their monetary payoffs conditional on being honest.\footnote{When a message is both optimal and precise, namely $\{10\}$, we count it as an optimal message. Likewise, when a message is both optimal and contains two numbers, like $\{8, 10\}$, we label it `Optimal.' We classify sub-optimal messages using two numbers as `Pair.'} Yet there remains a noticeable number of precise truth-tellers who reported drawing a number below 10, which contradicts our hypothesis. This may suggest the possibility of other motivation for truth-telling that is not captured by our model.

      \setcounter{RS}{2}
    \begin{subtheorem}{RS}
        \setcounter{RS}{1}
        \begin{RS}
            In the A-UR environment,
            \begin{enumerate}
                \item 46.4\% (13 of 28) and 14.3\% (4 of 28) of vague messages are pseudo-optimal and pairs, respectively; and
                \item 32\% (9 of 28) of vague messages take the form of an interval.
            \end{enumerate}
        \end{RS}
    \end{subtheorem}

\begin{figure}[h!]
        \centering
        \includegraphics[width=0.95\textwidth]{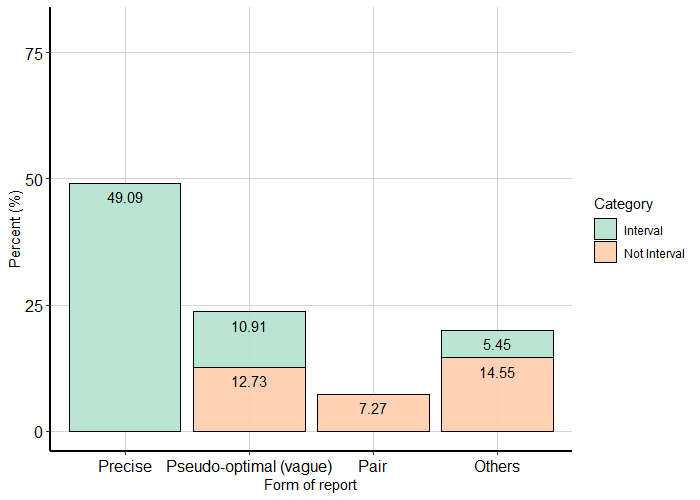}
        \caption{Message types used in A-UR treatments}
        \label{fig:report_form_A-UR}
    \end{figure}

Figure \ref{fig:report_form_A-UR} displays the various message types employed in A-UR treatments. In summary, 49.09\% (27 of 55) of participants utilized precise messages, while 50.91\% (28 of 55) favored vague messages.
In this treatment, we are unable to directly observe the true observations for each subject. To analyze subjects' behavior, we have developed a classification system for vague messages, dividing them into three categories: 'Pseudo-optimal,' 'Pair,' and 'Others.' We classify optimal-appearing messages as 'Pseudo-optimal,' an approach that parallels the optimal vague message (OVM) as defined in in Definition \ref{def:ovm}. To further refine our analysis, we use the minimum reported numbers in the messages as pseudo-true-observations. For example, a report consisting of $\{6, 9, 10\}$ is considered pseudo-optimal because this message would maximize the expected payoff if the true observation was 6. Conversely, a report $\{6, 7, 8, 9, 10\}$ would not be classified as pseudo-optimal.\footnote{When a message is both pseudo-optimal and pair, like $\{8, 10\}$, we count it as `Pseudo-optimal,' similar to how we classified vague messages in the AO-UR treatments. Only sub-optimal messages using two numbers are labeled `Pair.'}. Among the 28 vague messages, 13 were classified as pseudo-optimal (46.4\%) and 4 as pairs (14.3\%). Additionally, 9 out of the 28 messages (32.1\%) took the form of an interval.

         \setcounter{RS}{2}
    \begin{subtheorem}{RS}
        \setcounter{RS}{2}
        \begin{RS}
            In the NA-UR environment,
            \begin{enumerate}
                \item 22.0\% (13 of 59) and 13.6\% (8 of 59) of vague messages are pseudo-optimal and pairs, respectively; and
                \item 49.2\% (29 of 59) of vague messages take the form of an interval.
            \end{enumerate}
        \end{RS}
    \end{subtheorem}

\begin{figure}[ht]
        \centering
        \includegraphics[width=0.95\textwidth]{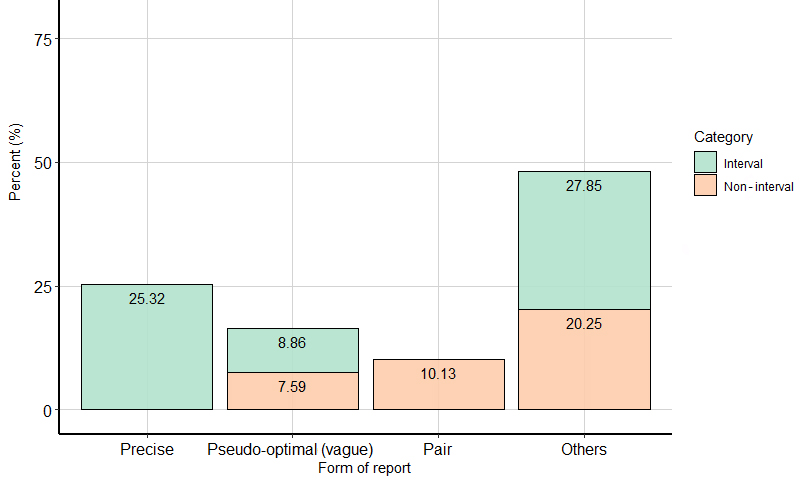}
        \caption{Message types used in NA-UR treatments}
        \label{fig:report_form_NA-UR}
    \end{figure}

    Figure \ref{fig:report_form_NA-UR} provides an overview of the message types employed in NA-UR treatments. In summary, 25.32\% (20 of 79) of participants utilized precise messages, while the majority 74.68\% (59 of 79) favored vague messages. Within the vague messages category, 13 were classified as pseudo-optimal (22.0\%) and 8 as pairs (13.6\%). Additionally, a considerable portion, 29 of them (49.2\%), took the form of an interval. 

    When comparing participant behavior between non-anonymous and anonymous treatments, an interesting observation is that a higher percentage of participants opted for vague messages in the non-anonymous setting.  This difference might reflect the fact that some  maximal liars in the anonymous case lie by reporting 10 in the absence of any concern with  social identity. Additionally, it is noteworthy that vague messages in the non-anonymous treatments were longer (with an average of 2.81 numbers reported) compared to those in the anonymous treatments (averaging 2.10 and 2.05 numbers reported for the AO and A treatments, respectively). Table \ref{tab:basic_result} displays the results of the difference in means tests. The differences are significant at the 1\% significance level, which could indicate the influence of vague messages on the external cost of lying.

      It is worth noting that, in comparison to the results observed in the anonymous treatments, as shown in Figures \ref{fig:report_form_AO-UR} and  \ref{fig:report_form_A-UR}, participants used substantially fewer precise and pair-type messages in the NA-UR treatment (\ref{fig:report_form_NA-UR}). Conditional on using vague messages, only 35.6\% are either pseudo-optimal or pairs in the NA-UR treatment. This ratio differs starkly from the 53.2\% and 60.7\% observed in the AO-UR and A-UR treatments, respectively. In addition, we see that the vast majority of the messages now fall into the `Others' category. Among the `Others' messages, half take the form of intervals, which was not the case in the anonymous treatments. In particular, 49.1\% of vague messages reported in the NA-UR treatment used intervals, while only 37.2\% and 32\% did so in the AO-UR and A-UR treatments, respectively. These findings suggest that subjects recognized that pseudo-optimal and pair messages are obvious and could negatively affect the audience's belief in an equilibrium\footnote{Note that if the receiver could correctly infer that the lowest number reported represents the true state, then the pseudo-optimal and pair messages should bear only the null social identity cost. If this were the case, participants in the NA-UR treatment should exploit these messages liberally to increase their expected payoffs. The fact that subjects did not do so implies that these messages are discounted in equilibrium, leading to a reduced social identity payoff. For instance, one might consider an extreme case like the one we proposed in Example \ref{ex:equilibrium_interval}, or a more general case where including higher numbers is typically discounted, yet including more numbers could mitigate the negative effect by rendering a better chance of being perceived as true.}.
      
\begin{figure}[ht]
        \centering
        \includegraphics[width=0.9\textwidth]{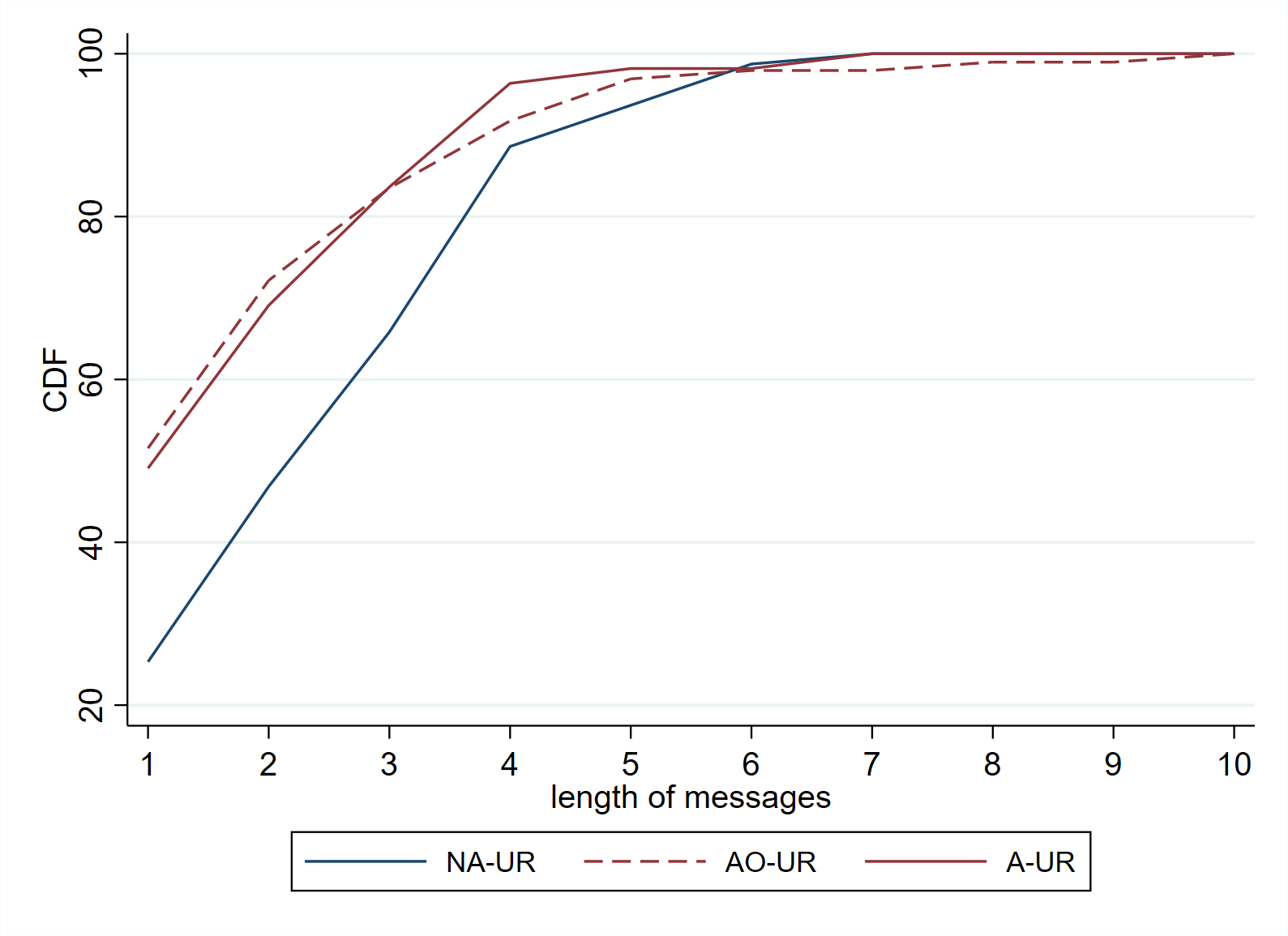}
        \caption{The cumulative distribution of lengths of vague messages in the 
        NA-UR, AO-UR and A-UR treatments}
        \label{fig:vague_length}
    \end{figure}

    The distribution of the lengths of messages also indicates that there is a significant difference in reporting patterns between the anonymous and non-anonymous treatments. In Figure \ref{fig:vague_length} we report the cumulative distribution of message lengths for the NA-UR, AO-UR and A-UR treatments. The figure  shows that subjects use vague and longer messages more often in the NA-UR treatment\footnote{The Kolmogorov-Smirnov test results reveal significant differences in the equality of the CDFs between non-anonymous and anonymous treatments. When comparing NA-UR to AO-UR, the null hypothesis of equal CDFs is rejected at the 1\% significance level (p-value = 0.005), while for the comparison between NA-UR and A-UR, the null hypothesis is rejected at the 10\% significance level (p-value = 0.051).}. Together with the frequent use of interval messages in the non-anonymous treatment, we interpret this pattern as indirect evidence of the impact of vagueness on the external cost of lying.

    \begin{RS}Comparison between the A, AO and and NA environments
       \begin{enumerate}
           \item The average numbers reported in NA-R, AO-R and A-R are 7.000,  7.619 and 8.163, respectively;
           \item the average numbers reported in NA-UR, AO-UR and A-UR are 8.282, 8.132 and 8.490, respectively.
       \end{enumerate}
    \end{RS}
    
    When comparing the behavior between the NA and AO treatments, we did not find statistically significant differences between the average numbers reported in them. We did not reject the null hypotheses in our results under either restricted or unrestricted communication, with p-values of 0.185 and 0.562, respectively. On the other hand, when comparing the behavior between the NA and A treatments, we did find a statistically significant difference in the average reports under restricted communication (p-value = 0.017). The latter is consistent with Hypothesis \ref{hp:NA-R-A-R} prediction, which states that subjects earn higher monetary payoffs on average in the anonymous environment. One possible explanation of the discrepancy in the results is that the observability of the true state increases the social-image costs of lying. This finding is in line with \textcite{Gneezy2018}, \textcite{Abeler2019} and \textcite{fries2021observability}, who find that lying decreases when the observability of the random draw increases. Specifically, \textcite{fries2021observability} conducted an experiment in which participants had the opportunity to lie about the outcome of a random draw in a private or public setting. They found that the proportion of participants who lied decreased significantly when the draw was observable to others. This supports the idea that increased observability can increase social-image concerns and decrease lying behavior, as suggested by our own findings in the present study. 
    
    Yet, our findings differ from those of \textcite{Fischbacher2013} and \textcite{fries2021observability} regarding the effect of anonymity on average reports. While they conclude that anonymity has no significant effect on the average reporting, we find that anonymity alone does have a significant impact on the reported numbers under restricted communication. This discrepancy may be attributed to differences in the experimental settings. 
    In the baseline treatment of \textcite{Fischbacher2013}, participants privately rolled a die and reported their roll and corresponding payoff on a screen, while the experimenter had the potential to match reported numbers with individual participants after the experiment. Conversely, in their anonymous treatment, participants anonymously deposited their remaining coins into a secondary envelope, preventing the experimenter from identifying who reported what number. Similarly, in the study by \textcite{fries2021observability}, participants were in isolated cubicles, where they privately rolled a die, recorded the result, and took out corresponding earnings from an envelope. In the anonymous treatments, participants sealed their report and the remaining money in the envelope, dropped it into an exit box, and left the lab, maintaining anonymity and avoiding interaction with the experimenter. In our current study, we aimed to replicate these previous settings within an online experiment, expecting to obtain similar results. In our anonymous treatment, participants were asked to choose a screen name, which served as an equivalent to the usage of envelopes in previous studies. However, we observed a disparity in results, which might be attributed to the possibility that participants felt a higher sense of anonymity in our online environment, compared to the in-person setting of previous studies.
    
    It is important to note that we did not find a significant difference in reported numbers between the anonymous and non-anonymous treatments under unrestricted communication. This suggests that the effect of anonymity on reported numbers may depend on the level of communication restriction. Therefore, future research could aim to identify the specific contextual factors that may modulate the effect of anonymity on reported numbers, and to investigate the underlying mechanisms that drive these effects.

\section{Concluding Remarks}

We contribute to the literature by connecting the study of lying behavior with investigations involving vague communication. Through our experiment, we examine the premise that conveying a vague yet truthful message carries a lower cost in terms of lying compared to delivering a precise maximal lie. This paper provides further insights into the costs associated with lying, particularly in relation to concerns regarding social identity. Moreover, our research extends the paradigm of lying experiments introduced by \cite{Fischbacher2013}. This extension involves integrating vague communication into the existing framework by employing set-valued messages. By doing so, our study advances our understanding of how the message space in communication influences an agent's reporting decision when lying carries a cost. 

Our study yields several findings. First, we observe that most subjects use vague messages and report higher values on average when when given the option to use vague messages.  Specifically, in sessions where social identity concerns are not pertinent, we find that individuals tend to lie less frequently and report higher values when provided with the opportunity to employ vague messages.  This finding suggests that participants strategically employ vagueness to maintain consistency with the truth while simultaneously leveraging the imprecision to their advantage and supports the key conjecture that a vague yet truthful message reduces the internal cost of lying.
This finding sheds new light on our understanding of lying aversion, suggesting that a restricted message space could be a source of the observed abstention from monetary-payoff maximization in previous experiments reported in the literature. In other words, the availability of using vague messages allows participants to maintain their integrity by incorporating true information in their messages while incurring lower monetary costs. In addition, we show that subjects are more inclined to use vague messages when they are motivated by social identity concerns. Furthermore, we find that when the social identity concern is relevant, subjects employ vague messages in longer and more sophisticated forms. This finding suggests that the utilization of vague messages offers individuals an alternative approach to maintain their social identity. In other words, when participants are provided with the option to use vague messages as a means to convey their honesty to an audience, they no longer need to sacrifice their monetary payoffs significantly in order to provide a credible signal. 

Our finding is analogous to the ``moral wiggle room'' mentioned in  \textcite{Dana2007}, where dictators care only whether they maintain an image of fairness, but without having a significant effect on  their partners' payoffs. In our experiment, most subjects exhibit no hesitation in increasing their monetary payoffs as long as their messages can remain even remotely truthful. The presence of precise truth-tellers in the anonymous environment, while small, suggests however that the moral-wiggle-room argument alone may not paint the whole picture. The decomposition of the aversion to monetary-payoff maximization in our experimental design calls for a new perspective on misleading behavior by showing that the observed aversion in many individuals is independent of the consequences of their message choices. It is possible that another motivation for truth-telling, such as a concern with projecting a well-intentioned self-image. That is, although subjects might have understood that there was no external observer to judge their behavior, their moral standards may combine the honesty achieved by reporting true observations with the uprightness of reporting the most accurate messages.

\pagebreak
\begin{sloppypar}
\printbibliography    
\end{sloppypar}

\pagebreak

\appendix
\appendixpage
\section{Proofs}
Let us first begin with restating the previously known results about the NA-R environment. The main intuition is that some agents lie when they see a small number, while some others always report the truthful message. Thus, when a reported number is small, it is safe to believe that the agent is being honest. The following equilibrium results establish a baseline for comparison with the other environments.
            \begin{theoremEnd}[proof at the end, text link= ]{lemma}[KS LM 1]
                In any equilibrium in the NA-R environment, there exists a strict positive probability that agents lie.
                \label{lm:NA-R_lie}
            \end{theoremEnd}
            \begin{proof}
                Suppose there exists an equilibrium in the NA-R environment where no agent lies and the audience believes  $P(\text{$j$ is truthful})=1$ for any report $j$. The utility of an arbitrary agent with the true observation of $i<N$ and the type $t$ is
                \begin{equation*}
                    U(i,j,t) = j - \mathds{1}(i \neq j) \left[ t + c(i,j) \right] + \gamma \cdot 1.
                \end{equation*}
                when the agent reports some $j$.

                As the social identity is constant under the audience's belief that everyone is truth-telling, the agent is better off by reporting $i+1$ instead of $i$ when
                \begin{equation*}
                    U(i,i+1,t) - U(i, i, t) = \left((i+1) - i\right) -  (t + c(i, i+1)) >0.
                \end{equation*}

                Because we assumed $c(i, i+1)<1$, there exists some $0 < t < 1- c(i, i+1)$.

            \end{proof}

            \begin{theoremEnd}[proof at the end, text link= ]{lemma}[KS PR 4; GKS PR 2]
                In any equilibrium in the NA-R environment, no agent underreports.
                \label{lm:NA-R_no-under}
            \end{theoremEnd}
            \begin{proof}
                Suppose there exists an equilibrium where an agent lies by reporting a number $j$ below their true observation $i$. Then it must be the case $\rho(j)>\rho(i)$.

                It also follows that reporting $j$ must yield a larger utility than reporting $i$:
                \begin{equation*}
                    (j-i)  + \gamma(\rho(j)- \rho(i)) \geq t + c(i,j).
                \end{equation*}

                We complete the proof by showing that $\rho(i)=1$; that is, if any agent choose to report $i$, then it must be the case that the report is truthful.

                Take another agent with the intrinsic aversion type $t$ who observes some $i^\prime \neq i$. We claim that $U(i^{\prime}, j, t^{\prime}) > U(i^{\prime}, i, t)$:
                \begin{align*}
                    (j-i) + [c(i^{\prime},i) - c(i^{\prime},j)] + \gamma(\rho(j) - \rho(i)) & \geq  c(i^{\prime},i) - c(i^{\prime},j) + t + c(i,j) >0
                \end{align*}
                because of the triangular inequality assumption: $c(i, i^{\prime}) + c(i^{\prime}, j) \geq c(i,j)$.
                As the choice of this agent is arbitrary, this is the case for all agents who do not observe $i$ never lies by reporting $i$; in turn, this implies $\rho(i)=1$, a contradiction.
            \end{proof}

            \begin{theoremEnd}[proof at the end, text link= ]{lemma}[KS THM 1; GKS PR 5]
				In any equilibrium in the NA-R environment,
				    \begin{enumerate}[i.]
				        \item there exists a threshold $1<l^{*}<N$ such that
				$$
					\forall j \geq l^{*} \; \exists i \neq j , t\in \mathbb{R} \; \text{ s.t. } \sigma^{j}_{it} > 0 \; \text{ and }
				$$
				$$
					\forall j < l^{*}, i\in \Omega, t \in \mathbb{R} \; \sigma^{j}_{it}=0;
				$$
				        \item all agents who observe a value above the threshold report their observed value truthfully.
				    \end{enumerate}
				\label{lm:NA-R_threshold}
			\end{theoremEnd}
			\begin{proof}
				Let $L^{\Omega}_{P} \subseteq M^{\Omega}_{P}$ be the set of messages that liars use to lie with positive probability. Let $l^{*} = \min L^{\Omega}_{P}$. By Lemma \ref{lm:NA-R_lie}, $L^{\Omega}_{P}$ is nonempty and $l^{*}$ is well-defined. Also, by the no-underreporting condition, we can deduce that $\rho(1)=1$ and $1<l^{*}$.

				We now show $L^{\Omega}_{P} = \{l^{*}, l^{*}+1, \dotsc, N\}$ by contradiction: suppose there exists some elements of $\Omega$ greater than $l^{*}$ which is not an element of $L^{\Omega}_{P}$. Let $n$ be the minimum of such elements, so that $n-1 \in L^{\Omega}_{P}$. As $\rho(n)=1$, we can see that any agent who lies by reporting $n-1$ is strictly better off by reporting $n$ instead:
				\begin{align*}
					U(i, n, t)- U(i, n-1 , t) & = (n - (n-1)) - (c(i,n) - c(i,n-1)) + \gamma(\rho(n) - \rho(n-1)) \\
					& > c(n-1, n) - (c(i,n) - c(i,n-1))  + \gamma(\rho(n) - \rho(n-1)) \geq 0.
				\end{align*}

				Now it remains to show that the agents who observe $i \in L^{\Omega}_{P}$ reports truthfully. Suppose there exists $j, j^{\prime} \in L^{\Omega}_{P}$ such that some agent observing $j$ instead chooses to report $j^{\prime}$. That is,
				\begin{align*}
					U(j, j^{\prime}, t)- U(j, j, t) & = (j^{\prime}-j) - (t + c(j, j^{\prime})) + \gamma(\rho(j^{\prime}) - \rho(j)) \geq 0
				\end{align*}
				for any intrinsic aversion type $t$.
				This implies that any agent who lies by reporting $j$ is strictly better off by reporting $j^{\prime}$ instead:
				\begin{align*}
					U(i, j^{\prime}, t)- U(i, j, t) & \geq (j^{\prime}-j) + (c(i, j) - c(i, j^{\prime})) + \gamma(\rho(j^{\prime}) - \rho(j)) \\
					& \geq t + c(j, j^{\prime}) + (c(i, j) - c(i, j^{\prime})) > 0
				\end{align*}
				for any intrinsic aversion type $t$ and any true observation $i \neq j$. Therefore, all agents who observe $i \in L^{\Omega}_{P}$ reports truthfully.
			\end{proof}

			Based on the three results, we bring a simple comparison of the behavior of an agent in the NA-R environment to that in the A-R environment. A thought experiment of choosing an agent and comparing their behavior in the two environments easily leads to a conjecture that the absence of the social identity concern should only facilitate more lies. Applying these observation to the unrestricted communication leads to Proposition \ref{pr:use_vague}. We also obtain Lemma \ref{lm:NA-R_then-A-R} from these three lemmas, which we will use to prove Proposition \ref{pr:NA-R-A-R}.
			
\printProofs

\section{Gender Differences}
\subsection{Non-anonymous Environment}
In the non-anonymous environment, 28 males and 51 females attended the experiments. We do not find statistically significant differences between males and females neither on their reports nor on the length of vague messages.

\begin{table}[!htbp] \centering 
  \caption{Gender differences on Reports in the Non-anonymous Environment} 
  \label{tab:gender_NA} 
  \small
\begin{tabular}{@{\extracolsep{5pt}}lccc} 
\\[-1.8ex]\hline 
\hline \\[-1.8ex] 
\\[-1.8ex] & Restricted & Unrestricted & Length \\ 
\hline \\[-1.8ex] 
 Gender differences & 0.055 & $-$0.159 & 0.294 \\ 
  & (0.740) & (0.290) & (0.390) \\ 
 \hline \\[-1.8ex] 
Observations & 79 & 79 & 79 \\ 
\hline 
\hline \\[-1.8ex] 
\end{tabular} 
\end{table} 

Under the unrestricted communication, male and female subjects are almost equally likely to send a precise message. Men are more likely to send a pseudo-optimal vague message, while women are slightly more likely to send a message with contains a pair or an interval.

 \begin{figure}[ht]
        \centering
        \includegraphics[width=13cm,height=8cm]{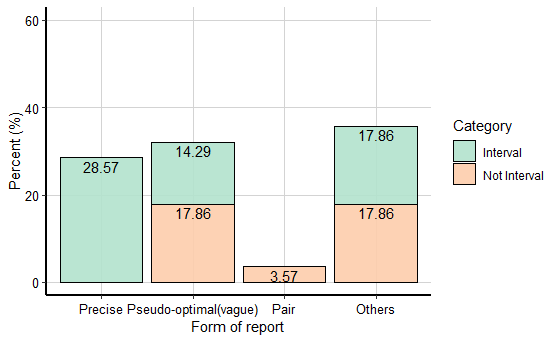}
        \caption{Message types male subjects used in NA-UR treatments}
        \label{fig:male_report_form_NA-UR}
    \end{figure}

 \begin{figure}[ht]
        \centering
        \includegraphics[width=13cm,height=8cm]{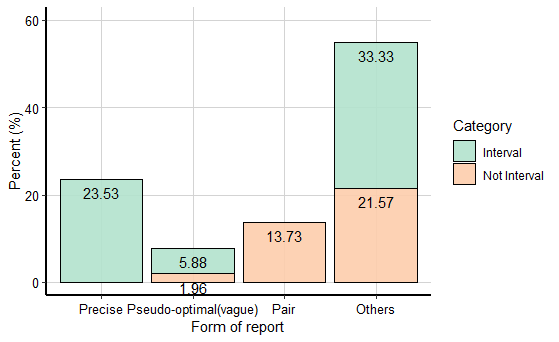}
        \caption{Message types female subjects used in NA-UR treatments}
        \label{fig:female_report_form_NA-UR}
    \end{figure}
    
\FloatBarrier

\subsection{Anonymous Environment}

 We used only a subset of the anonymous and observable sessions to analyze the gender effect as we could not identify the participants' gender in principle. We ran four single sexed sessions (two male-only sessions and two female-only sessions) to create a gender dummy variable without interfering the anonymous property of the experiment design.
 
Under the restricted communication, we do not see statistically significant differences between males and females neither on their reports nor on the length of vague messages. 

\begin{table}[!htbp] \centering 
  \caption{Gender differences on Reports in the Anonymous Environment} 
  \label{tab:gender_A} 
  \small
\begin{tabular}{@{\extracolsep{5pt}}lccc} 
\\[-1.8ex]\hline 
\hline \\[-1.8ex] 

\\[-1.8ex] & Restricted & Unrestricted & Length \\ 

\hline \\[-1.8ex] 
 Gender differences & 0.103 & $-$0.606 & $-$0.487 \\ 
  & (0.715) & (0.594) & (0.428) \\ 

 \hline \\[-1.8ex] 
Observations & 64 & 64 & 64 \\ 
\hline 
\hline \\[-1.8ex] 
\end{tabular} 
\end{table} 
 
\begin{figure}[ht]
        \centering
        \includegraphics[width=11cm,height=8cm]{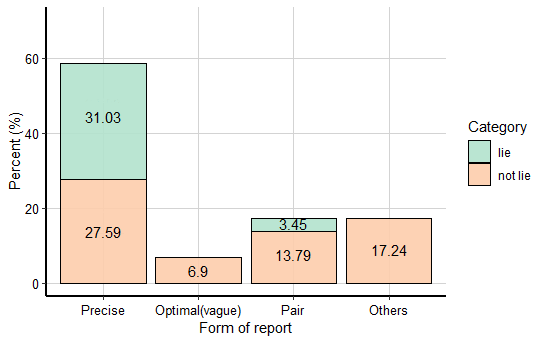}
        \caption{Message types male subjects used in AO-UR treatments}
        \label{fig:male_report_form_A-UR}
    \end{figure}

\begin{figure}[ht]
        \centering
        \includegraphics[width=11cm,height=8cm]{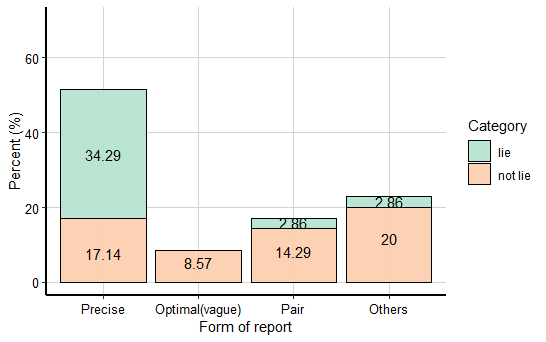}
        \caption{Message types female subjects used in AO-UR treatments}
        \label{fig:female_report_form_A-UR}
    \end{figure}

The results show that, 34.29 \% of males and 31.03\% of females lied in the non-anonymous treatment. Therefore, we do not find substantial difference in the lying behavior between them. In addition, we do not see significant differences in the choice of type and length of vague messages, as shown by Figures \ref{fig:male_report_form_A-UR} and \ref{fig:female_report_form_A-UR}.

\FloatBarrier

\section{Experimental Instructions}

\subsection{Instructions for Non-Anonymous-Non-observable session (Zoom session; video)}
Welcome to the experiment! Thank you very much for participating today. Before we start, I will go over the roster to take attendance to make sure I have everyone registered for the session. During the attendance please turn your video on.

I will now walk you through the instructions for the experiment. We need your full attention during the experiment. If you have trouble with hearing the audio or seeing the shared screen, please let me know. If you have any questions during the instructions, please use the hand-raising feature of Zoom and your question will be answered out loud, so everyone can hear. 

The experiment you will be participating in today is an experiment in individual decision making. At the end of the experiment, you will be paid an Amazon gift card. You will receive the show-up fee of \$2 for completing the experiment, with the additional amount that depends on your decisions and on chance. The details of the payment will be described later. All instructions and descriptions that you will be given in the experiment are accurate and true. In accordance with the policy of this lab, at no point will we attempt to deceive you in any way.

This is a screenshot of the first page in the main experiment. At the end of the instructions, I will provide the link to the experiment using the chat window. Please copy and paste the link to your browser. 
The first screen will ask your identification information – your first and last name and your student ID. 

After you enter your information, you will proceed to the next screen and take a short quiz to ensure your understanding of the procedures. You will be able to repeat the quiz if you make mistakes. You will have three chances to attempt the quiz. If you fail to get all questions correct after three attempts, you may not participate in today’s experiment. In such case, you will only receive the show-up fee for today’s experiment.

The main part of the experiment consists of two STAGES after the preliminary quiz. In each STAGE you will observe a number that we ask you to remember and later report to us. The number you report to us determines how much money you will be paid. 
At the end of the experiment, a confirmation screen will summarize today’s experiment and provide the information to retrieve your payment. 
I will give you more details about the observation, reporting, and payment processes.

The observation process is identical in both STAGES. At the beginning of both STAGES, we will provide a link to Google page that randomly generates a number between 1 and 10. The probabilities are equal across the numbers; that is, each number is chosen with the same probability of one tenth. 
We ask you to open the link, remember the number, close the Google page, and report the number on the next screen. This is a screenshot of an example Google page. You will see that this is a search result for ‘random number between 1 and 10’, and the page displays a randomly generated number that matches the search phrase.
Do not click on the ‘generate’ button on the Google page, because the number you see is already randomly generated. Any additional generation only distorts the statistical properties of the experiment.
I will now demonstrate how this Google page works. You will find this link during the experiment, and this is equivalent of opening a google page and typing in ``random number between 1 and 10.'' When you open the link, a new window pops up. As you can see, this number is already randomly generated and you should not generate the number again.

The way you can report differs between STAGES. 
In one STAGE, you are allowed to select one number after you observe the randomly generated number. 
This is a screenshot of the experiment stage where you can select one number. By one number, we mean that you may click only one number on the screen. We will interpret this selection as a statement that the number you observed is the number you selected, and we will regard this number as your report.
The number you report determines how much money you will be paid. You will be paid the equivalent in dollars to the number you report divided by 2. In other words, if you report ``1'', you receive 50 cents. If you report ``2'', you receive \$1, if you report ``3'', you receive \$1.50 and so on. A confirmation screen after your report will help you review your selection and the corresponding payment.
In another STAGE, you are allowed to select a set of numbers after you observe the randomly generated number. 
This is a screenshot of the experiment stage where you can select a set of numbers. By a set of numbers, we mean that you may click multiple numbers on the screen. For instance, you may choose to click on four numbers, one number, two numbers, or even all ten numbers.
If you select multiple numbers, we will interpret it as a statement that the number you observed is one of the numbers you selected. If you select a single number, we will interpret it as a statement that the number you observed is the number you selected. 
After you submit your selection, the computer will randomly choose one number from the set of numbers you selected. We will regard this randomly chosen number as your report. Again, the number you report determines how much money you will be paid. 
You will be paid the equivalent in dollars to the randomly chosen number from the set of numbers you selected divided by 2. If the randomly chosen number is ``1'', you receive 50 cents. If ``2'', you receive \$1, if ``3'', you receive \$1.50 and so on. A confirmation screen after your report will help you review your selection and the corresponding payment.

The order of the two STAGES is randomly determined. In other words, it is equally likely that you either participate in the STAGE allowing a single number first and then participate in the STAGE allowing a set of numbers or participate in the STAGE allowing a set of numbers first and then participate in the STAGE allowing a single number. 
In any case, you will play each STAGE only once. 

After the completion of both STAGES, a final review screen will summarize today’s experiment. This is a screenshot of the review screen. The last screen will ask your email address to receive the Amazon Gift Code of the amount that corresponds to your responses. 
We will directly send you Amazon gift code to the email address you provide. Please allow us a few hours after the completion of the experiment to validate the data and send the email.

Washington University in St. Louis recommends student subjects to report their taxpayer identification information for tax purposes. If you are an international student and do not have the taxpayer identification information, please indicate so by entering ‘Foreign’ in the form. If you do not have or do not wish to provide the identification information, please indicate that you would like to opt out by entering 'Refuse' in the form.

This is the end of instructions. If you have any questions, please raise your hand. Otherwise, I will provide the link via the chat window. Please copy and paste the link to your browser and start the experiment.

\subsection{Instructions for Anonymous-Observable session (Zoom session; no video)}
Welcome to the experiment! Thank you very much for participating today, and I will first walk you through the instructions for the experiment.

We need your full attention during the experiment. If you have trouble with hearing the audio or seeing the shared screen, please let me know. Do not turn the video on during the experiment. If you have any questions during the instructions, please raise hand so that I can unmute you. Your question will be answered out loud, so everyone can hear.

The experiment you will be participating in today is an experiment in individual decision making. At the end of the experiment, you will be paid an Amazon gift card. You will receive the show-up fee of \$2 for completing the experiment, with the additional amount that depends on your decisions and on chance. The details of the compensation will be described later. All instructions and descriptions that you will be given in the experiment are accurate and true. In accordance with the policy of this lab, at no point will we attempt to deceive you in any way.

I would like to first point out that we want to ensure this experiment is conducted anonymously, meaning that we cannot connect the responses recorded in this experiment to any particular individual who participated in this research. Qualtrics, the survey platform we are using, provides an option for the researchers to not collect any personal information, such as IP address or geographic location of the participants, for anonymous surveys. 
Also, your response will be recorded with a SCREEN NAME. You will be asked to choose a screen name that is at least 8 characters in length using letters, numbers, and underscore. This SCREEN NAME is only used in data analysis and distributing your Amazon gift card after the experiment. We cannot and will not attempt to associate SCREEN NAMEs to any particular individual.

I will now describe the main features of the experiment. 
First, there is a short quiz to ensure your understanding of the procedures. You will be able to repeat the quiz if you make mistakes. You will have three chances to attempt the quiz. If you fail to get all questions correct after three attempts, you may not participate in the main experiment. Even in such a case, please remain connected to the Zoom session until everyone finishes, and you will receive the show-up fee for today’s experiment.
The main part of the experiment after the preliminary quiz consists of two STAGES. In each STAGE you will observe a number that we ask you to remember and later report to us. The number you report to us determines how much money you will be paid.
At the end of the experiment, a confirmation screen will summarize today’s experiment and provide the information to retrieve your payment. 
 I will give you more details about each step.
At the end of this instructions, we will first provide a link for the preliminary quiz using the chat window. This is a screenshot of the webpage. Please choose a screen name, and make sure you keep this screen name. 
After you successfully complete the preliminary quiz, you will be reminded of your screen name once again. We recommend you copy and paste the screen name. We cannot recover this information for you, and you will not be able to receive your compensation without the correct screen name.
Please wait while everyone else finishes the quiz. Once everyone finishes, we will provide another link for the main experiment using the chat window.
This is a screenshot of the first page of the main experiment. Make sure you use the same screen name you used in the preliminary quiz. You may not receive your compensation if the screen names do not match.
As previously mentioned, your main task today is to observe a number that we ask you to remember and later report to us. The observation process is identical in both STAGES. At the beginning of both STAGES, the computer will randomly draw a number between 1 to 10. The probabilities are equal across the numbers; that is, each number is chosen with the same probability of one-tenth. We ask you to remember the number and report on the next screen. 
However, the way you can report differs between STAGES. 
In one STAGE, you are allowed to select one number after you observe the draw.
This is a screenshot of the experiment stage where you can select one number. By one number, we mean that you may click only one number on the screen. We will interpret this selection as a statement that the number you observed is the number you selected, and we will regard this number as your report.
The number you report determines how much money you will be paid. You will be paid the equivalent in dollars to the number you report divided by 2. In other words, if you report ``1'', you receive 50 cents. If you report ``2'', you receive \$1, if you report ``3'', you receive \$1.50 and so on. A confirmation screen after your report will help you review your selection and the corresponding payment.
In another STAGE, you are allowed to select a set of numbers after you observe the draw. 
This is a screenshot of the experiment stage where you can select a set of numbers. By a set of numbers, we mean that you may click multiple numbers on the screen. For instance, you may choose to click on four numbers, one number, two numbers, or even all ten numbers.
If you select multiple numbers, we will interpret it as a statement that the number you observed is one of the numbers you selected. If you select a single number, we will interpret it as a statement that the number you observed is the number you selected. 
After you submit your selection, the computer will randomly choose one number from the set of numbers you selected. We will regard this randomly chosen number as your report.
Again, the number you report determines how much money you will be paid. 
You will be paid the equivalent in dollars to the randomly chosen number from the set of numbers you selected divided by 2. If the randomly chosen number is ``1'', you receive 50 cents. If ``2'', you receive \$1, if ``3'', you receive \$1.50 and so on. A confirmation screen after your report will help you review your selection and the corresponding payment.

The order of the two STAGES is randomly determined. In other words, it is equally likely that you first participate in the STAGE allowing only a single number and then participate in the STAGE allowing a set of numbers, or first participate in the STAGE allowing a set of numbers and then participate in the STAGE allowing only a single number. 
In any case, you will play each STAGE only once. 

After the completion of both STAGES, a final review screen will provide the information to receive your Amazon gift card code. This is a screenshot of the review screen.
This includes your SCREEN NAME you entered, the amount you will receive, and a randomly generated passcode. Because the experiment is anonymous, we have no means to recover this information for you. Please make sure you either print or take the screenshot of this page for your record, because it is very important when you retrieve your compensation.

In the final review screen, you will find a link of a google form. To receive your amazon gift card you have to copy and paste the link in another browser and then fill your personal information. The google form is created by a staff of the department of economics, who won't have access to the data of this experiment.The experimenter only has a list of SCREEN NAMES and the amount associated with them. The experimenter will never be able to access your personal information. This is a screenshot of the google form which you will have to fill to receive your amazon gift card. 

Washington University in St. Louis recommends student subjects to report their taxpayer identification information for tax purposes. If you are an international student and do not have the taxpayer identification information, please indicate so by entering ‘Foreign’ in the form. If you do not have or do not wish to provide the identification information, please indicate that you would like to opt out by entering 'Refuse' in the form.

We are sorry for the inconvenience that we are not able to email you with the gift code directly. This payment process is to ensure anonymity in this experiment, and we appreciate your understanding that the anonymity of the reports constitutes a crucial component of our research. 

This is the end of the instructions.  If you have any questions, please raise hand. Otherwise, I will provide the link via the chat window. Please copy and paste the link to your browser and start the experiment.

\subsection{Instructions for Anonymous-Non-observable session (Zoom session; no video)}

Welcome to the experiment! Thank you very much for participating today, and I will first walk you through the instructions for the experiment.
We need your full attention during the experiment. If you have trouble with hearing the audio or seeing the shared screen, please let me know. Do not turn the video on during the experiment. If you have any questions during the instructions, please raise hand so that I can unmute you. Your question will be answered out loud, so everyone can hear.
The experiment you will be participating in today is an experiment in individual decision making. At the end of the experiment, you will be paid an Amazon gift card. You will receive the show-up fee of \$2 for completing the experiment, with the additional amount that depends on your decisions and on chance. The details of the compensation will be described later. All instructions and descriptions that you will be given in the experiment are accurate and true. In accordance with the policy of this lab, at no point will we attempt to deceive you in any way.
I would like to first point out that we want to ensure this experiment is conducted anonymously, meaning that we cannot connect the responses recorded in this experiment to any particular individual who participated in this research. Qualtrics, the survey platform we are using, provides an option for the researchers to not collect any personal information, such as IP address or geographic location of the participants, for anonymous surveys. Also, your response will be recorded with a SCREEN NAME. You will be asked to choose a screen name that is at least 8 characters in length using letters, numbers, and underscore. This SCREEN NAME is only used in data analysis and distributing your Amazon gift card after the experiment. We cannot and will not attempt to associate SCREEN NAMEs to any particular individual.

I will now describe the main features of the experiment. First, there is a short quiz to ensure your understanding of the procedures. You will be able to repeat the quiz if you make mistakes. You will have three chances to attempt the quiz. If you fail to get all questions correct after three attempts, you may not participate in the main experiment. Even in such a case, please remain connected to the Zoom session until everyone finishes, and you will receive the show-up fee for today’s experiment. The main part of the experiment after the preliminary quiz consists of two STAGES. In each STAGE you will observe a number that we ask you to remember and later report to us. The number you report to us determines how much money you will be paid.
At the end of the experiment, a confirmation screen will summarize today’s experiment and provide the information to retrieve your payment. I will give you more details about each step. At the end of this instructions, we will first provide a link for the preliminary quiz using the chat window. This is a screenshot of the web page. Please choose a screen name, and make sure you keep this screen name. After you successfully complete the preliminary quiz, you will be reminded of your screen name once again. We recommend you copy and paste the screen name. We cannot recover this information for you, and you will not be able to receive your compensation without the correct screen name. Please wait while everyone else finishes the quiz. Once everyone finishes, we will provide another link for the main experiment using the chat window. This is a screenshot of the first page of the main experiment. Make sure you use the same screen name you used in the preliminary quiz. You may not receive your compensation if the screen names do not match. As previously mentioned, your main task today is to observe a number that we ask you to remember and later report to us. The observation process is identical in both STAGES. At the beginning of both STAGES, we will provide a link to Google page that randomly generates a number between 1 and 10. The probabilities are equal across the numbers; that is, each number is chosen with the same probability of one tenth. We ask you to open the link, remember the number, close the Google page, and report the number on the next screen. This is a screenshot of an example Google page. You will see that this is a search result for ‘random number between 1 and 10’, and the page displays a randomly generated number that matches the search phrase. Do not click on the ‘generate’ button on the Google page, because the number you see is already randomly generated. Any additional generation only distorts the statistical properties of the experiment. I will now demonstrate how this Google page works. You will find this link during the experiment, and this is equivalent of opening a google page and typing in “random number between 1 and 10.” When you open the link, a new window pops up. As you can see, this number is already randomly generated, and you should not generate the number again. However, the way you can report differs between STAGES. In one STAGE, you are allowed to select one number after you observe the draw.This is a screenshot of the experiment stage where you can select one number. By one number, we mean that you may click only one number on the screen. We will interpret this selection as a statement that the number you observed is the number you selected, and we will regard this number as your report. The number you report determines how much money you will be paid. You will be paid the equivalent in dollars to the number you report divided by 2. In other words, if you report “1”, you receive 50 cents. If you report “2”, you receive \$1, if you report “3”, you receive \$1.50 and so on. A confirmation screen after your report will help you review your selection and the corresponding payment. In another STAGE, you are allowed to select a set of numbers after you observe the draw. This is a screenshot of the experiment stage where you can select a set of numbers. By a set of numbers, we mean that you may click multiple numbers on the screen. For instance, you may choose to click on four numbers, one number, two numbers, or even all ten numbers. If you select multiple numbers, we will interpret it as a statement that the number you observed is one of the numbers you selected. If you select a single number, we will interpret it as a statement that the number you observed is the number you selected. After you submit your selection, the computer will randomly choose one number from the set of numbers you selected. We will regard this randomly chosen number as your report. Again, the number you report determines how much money you will be paid. You will be paid the equivalent in dollars to the randomly chosen number from the set of numbers you selected divided by 2. If the randomly chosen number is “1”, you receive 50 cents. If “2”, you receive \$1, if “3”, you receive \$1.50 and so on. A confirmation screen after your report will help you review your selection and the corresponding payment.
The order of the two STAGES is randomly determined. In other words, it is equally likely that you first participate in the STAGE allowing only a single number and then participate in the STAGE allowing a set of numbers, or first participate in the STAGE allowing a set of numbers and then participate in the STAGE allowing only a single number. In any case, you will play each STAGE only once. 

After the completion of both STAGES, a final review screen will provide the information to
receive your Amazon gift card code. This is a screenshot of the review screen. This includes your SCREEN NAME you entered and the amount you will receive. Because the experiment is anonymous, we have no means to recover this information for you. Please make sure you either print or take the screenshot of this page for your record, because it is very important when you retrieve your compensation.
In the final review screen, you will find a link of a google form. To receive your amazon
gift card you have to copy and paste the link in another browser and then fill your personal information. The google form is created by a staff of the department of economics, who won’t have access to the data of this experiment.The experimenter only has a list of SCREEN NAMES and the amount associated with them. The experimenter will never be able to access your personal information. This is a screenshot of the google form which you will have to fill to receive your amazon gift card.

Washington University in St. Louis recommends student subjects to report their taxpayer identification information for tax purposes. If you are an international student and do not have the taxpayer identification information, please indicate so by entering ‘Foreign’ in the form. If you do not have or do not wish to provide the identification information, please indicate that you would like to opt out by entering ’Refuse’ in the form.

We are sorry for the inconvenience that we are not able to email you with the gift code
directly. This payment process is to ensure anonymity in this experiment, and we appreciate
your understanding that the anonymity of the reports constitutes a crucial component of our
research.

This is the end of the instructions. If you have any questions, please raise hand. Otherwise, I will provide the link via the chat window. lease copy and paste the link to your browser and start the experiment.

\begin{figure}[ht]
	\centering
	\includegraphics[width=0.8\textwidth]{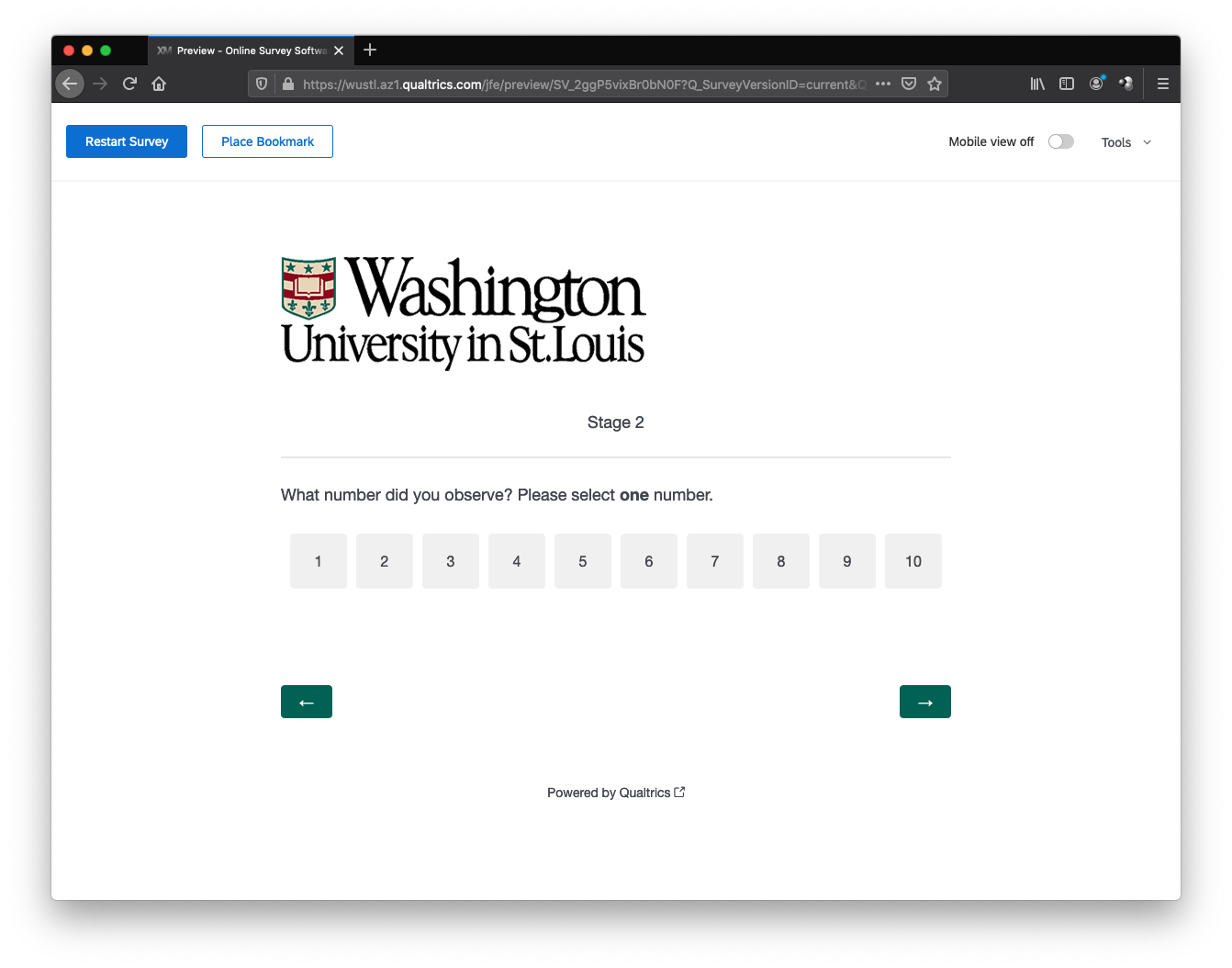}
	\caption{A screenshot of the experiment software displaying the precise-message stage}
	\label{fig:precise_screenshot}
\end{figure}
\begin{figure}[ht]
	\centering
	\includegraphics[width=0.8\textwidth]{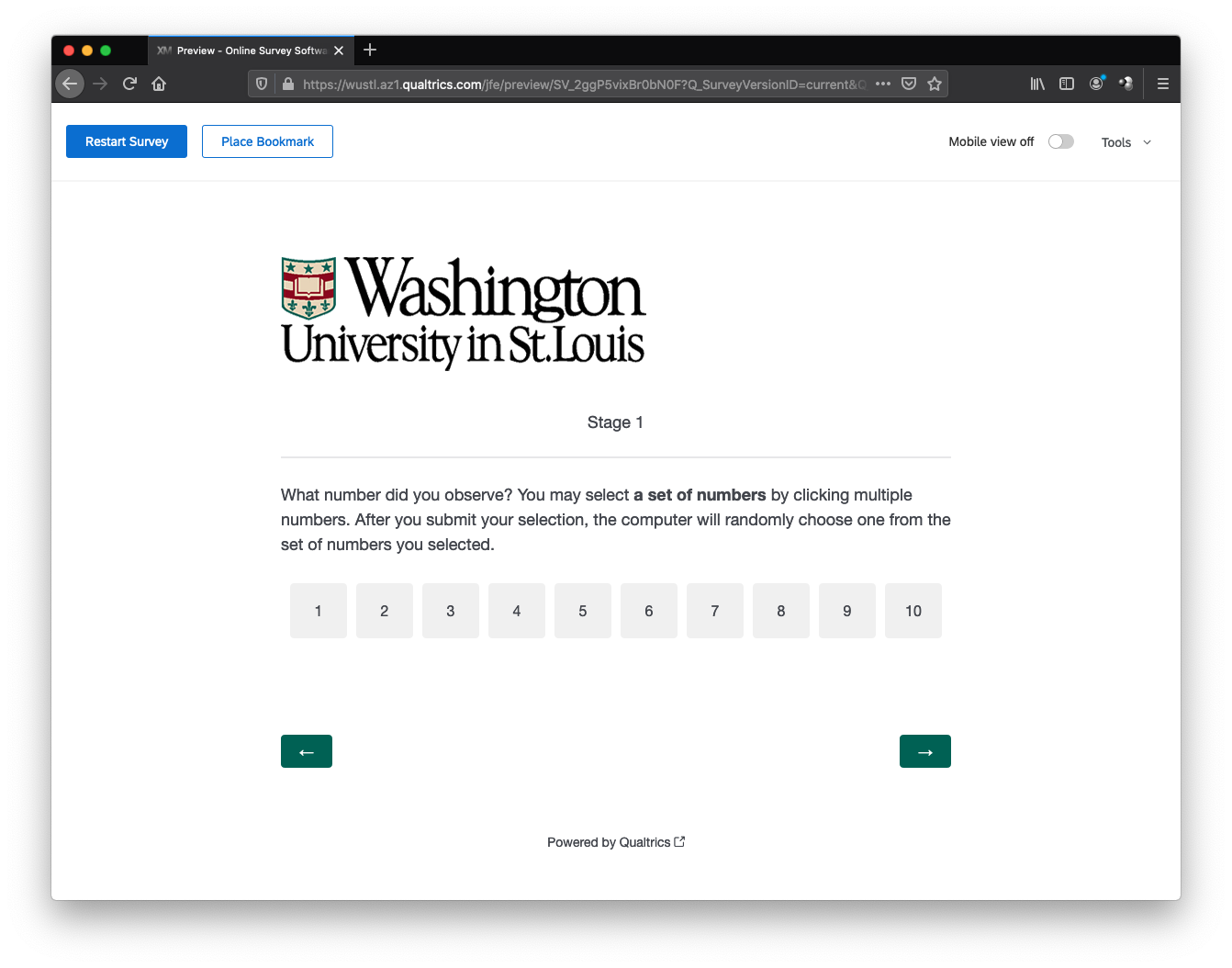}
	\caption{A screenshot of the experiment software displaying the vague-message stage}
	\label{fig:vague_screenshot}
\end{figure}

\begin{figure}
  \begin{subfigure}{0.3\textwidth}
    \centering
    \includegraphics[width=\linewidth]{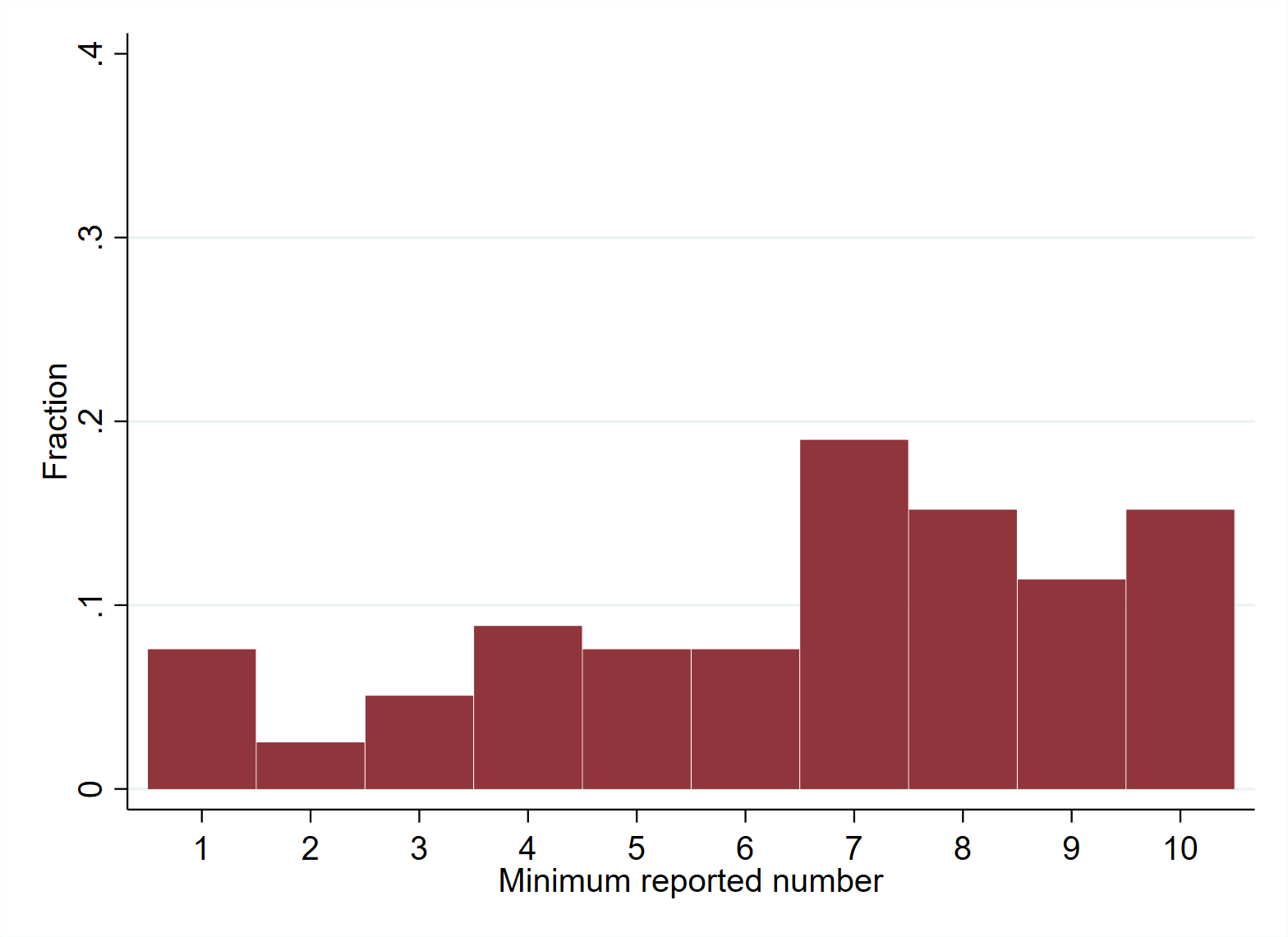}
    \phantomsubcaption %
    \label{fig:NA_R}
  \end{subfigure}%
  \hfill
  \begin{subfigure}{0.3\textwidth}
    \centering
    \includegraphics[width=\linewidth]{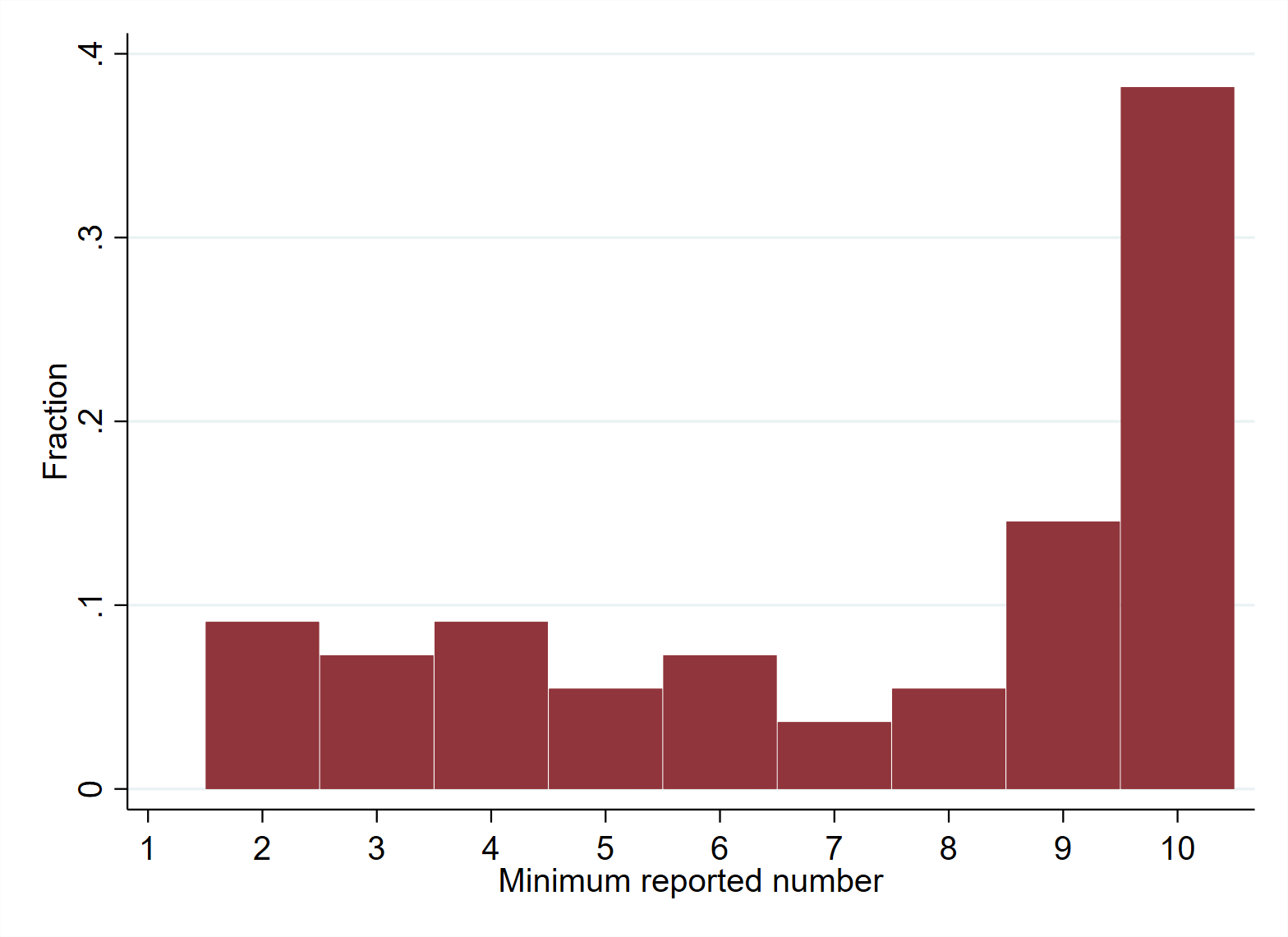}
    \phantomsubcaption %
    \label{fig:A_R}
  \end{subfigure}%
  \hfill
  \begin{subfigure}{0.3\textwidth}
    \centering
    \includegraphics[width=\linewidth]{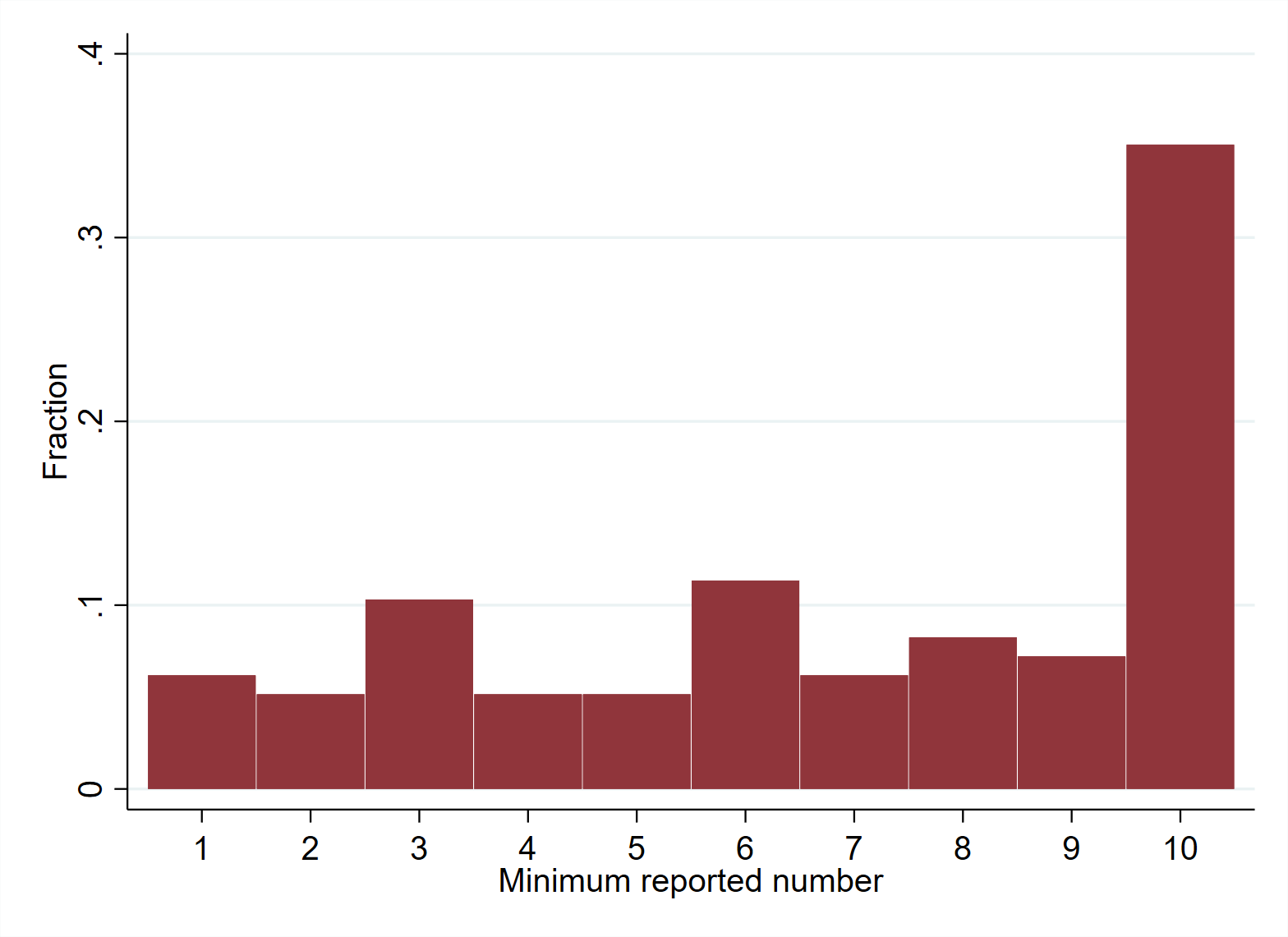}
    \phantomsubcaption %
    \label{fig:AO_R}
  \end{subfigure}
  \caption{Fractions of minimum reported numbers in NA-UR, A-UR, and AO-UR}
  \label{fig:minimum_unrestr}
\end{figure}

\begin{figure}
  \begin{subfigure}{0.3\textwidth}
    \centering
    \includegraphics[width=\linewidth]{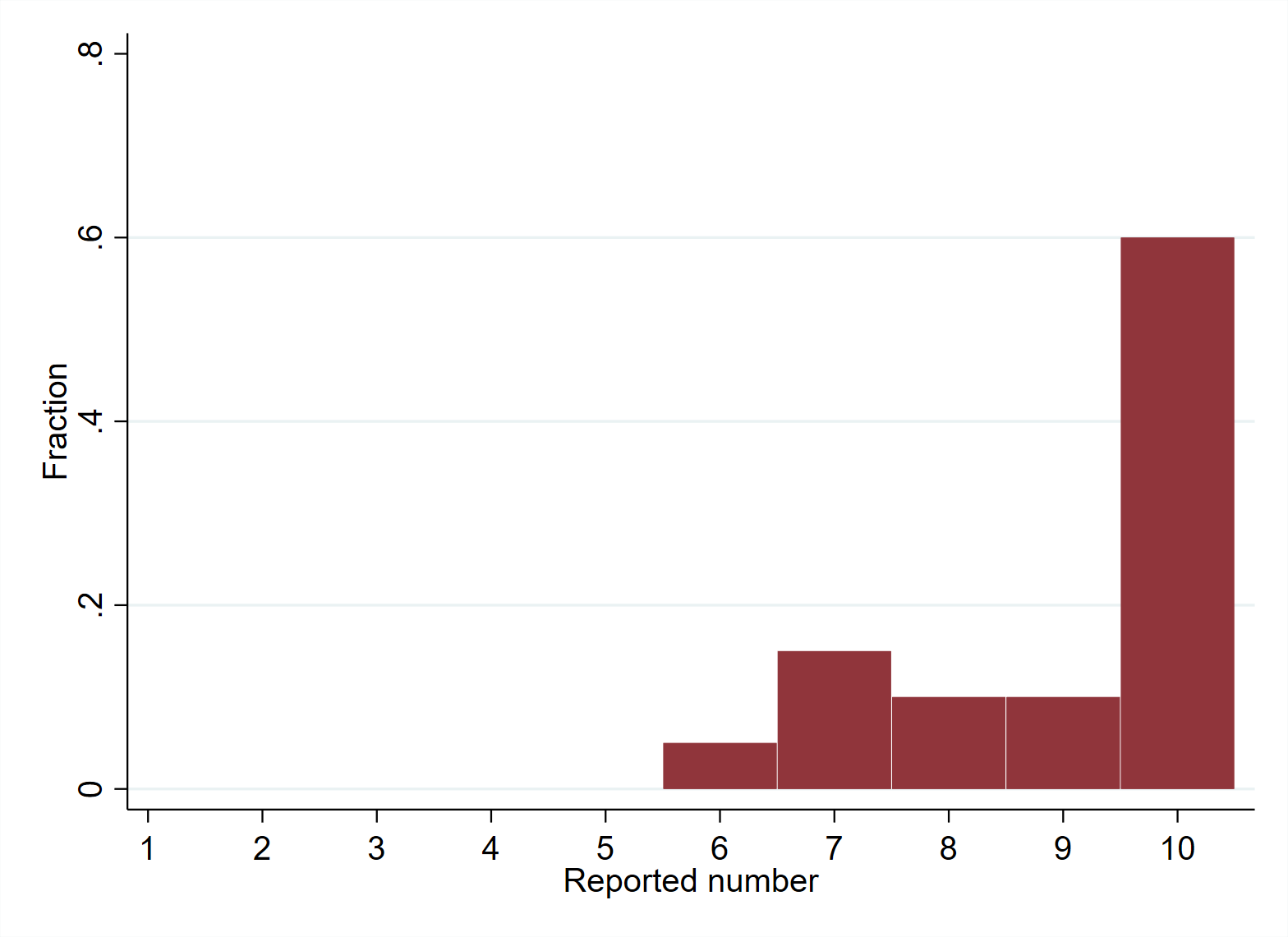}
    \phantomsubcaption %
    \label{fig:NA_R}
  \end{subfigure}%
  \hfill
  \begin{subfigure}{0.3\textwidth}
    \centering
    \includegraphics[width=\linewidth]{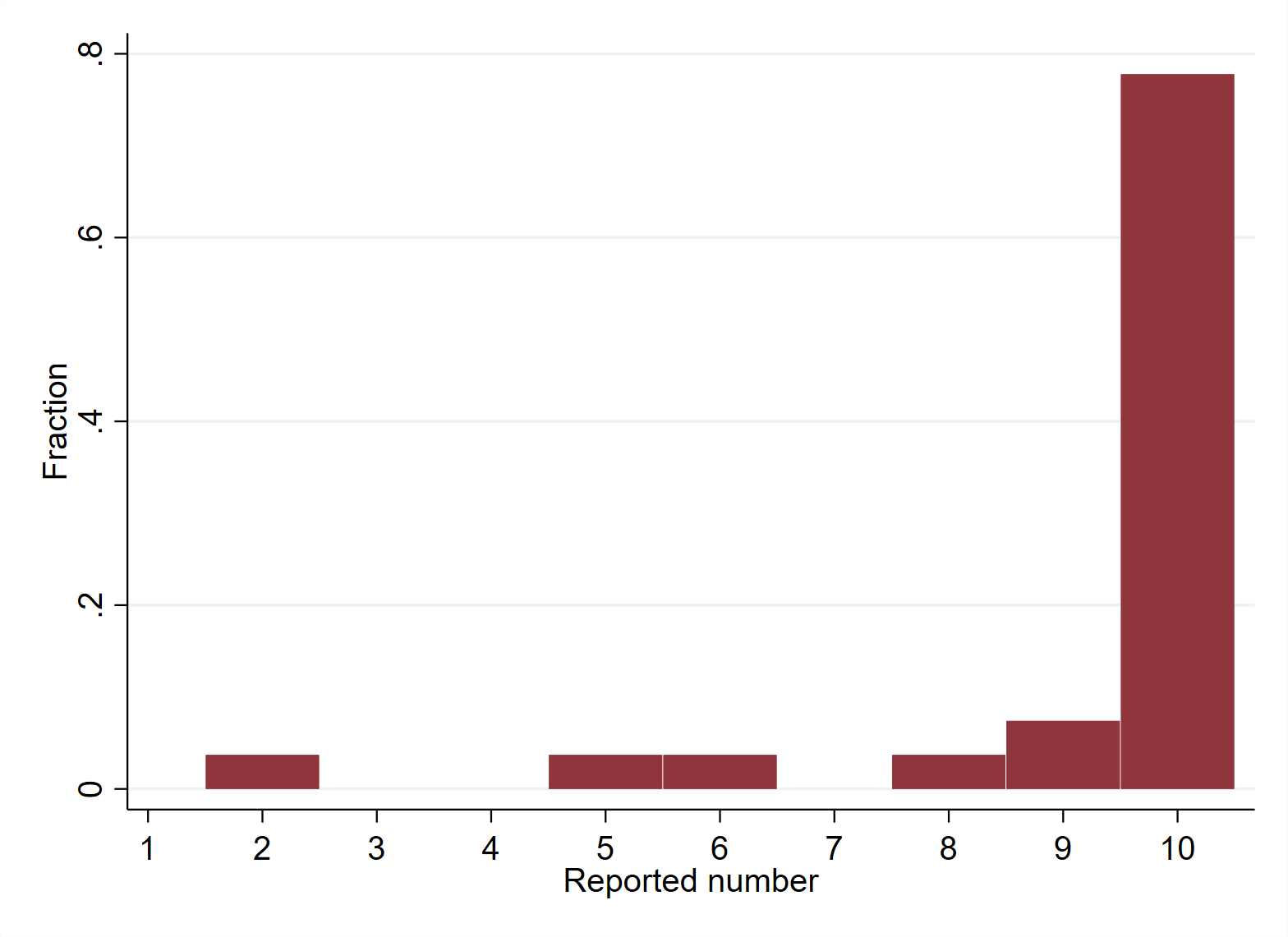}
    \phantomsubcaption %
    \label{fig:A_R}
  \end{subfigure}%
  \hfill
  \begin{subfigure}{0.3\textwidth}
    \centering
    \includegraphics[width=\linewidth]{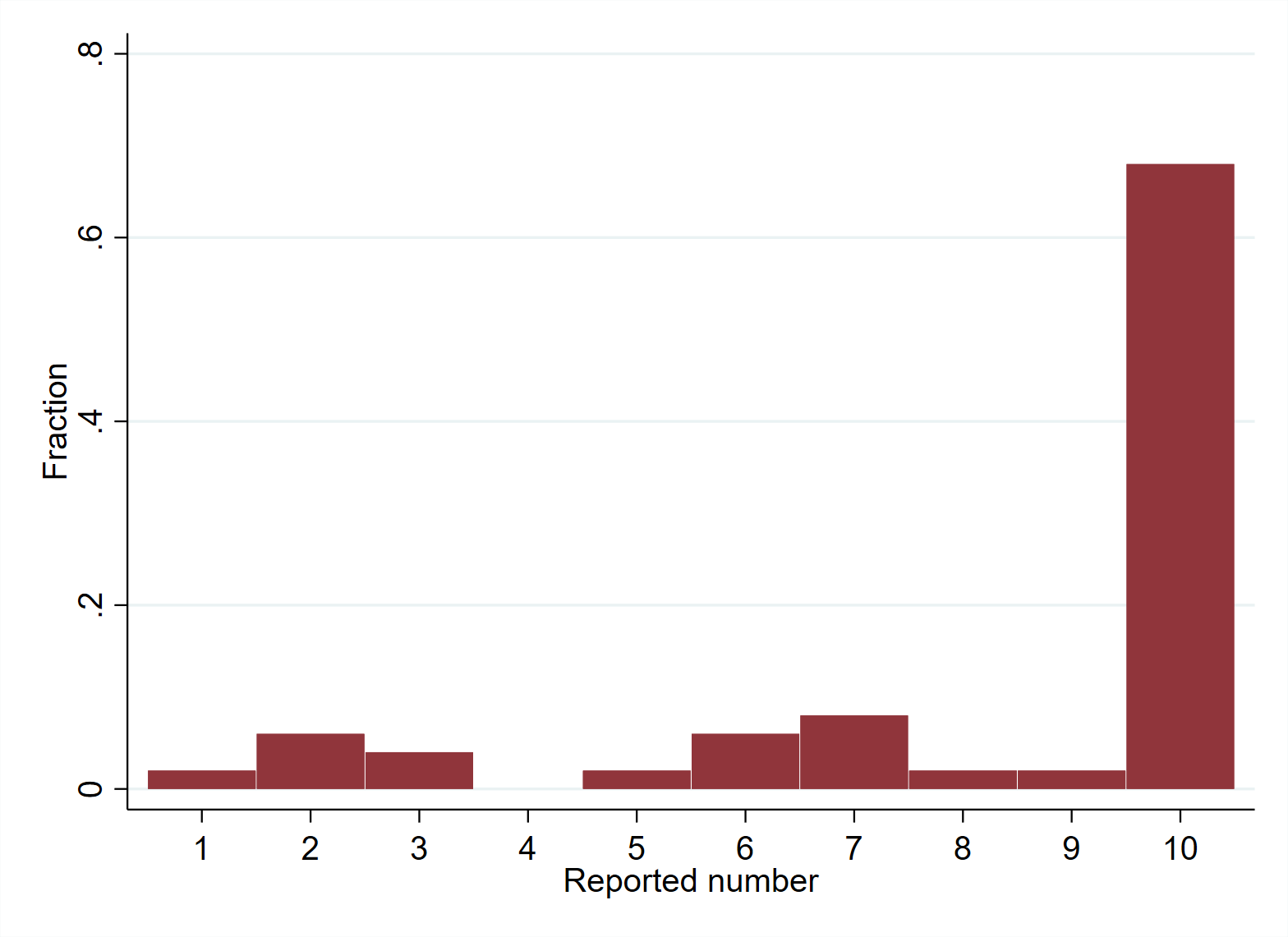}
    \phantomsubcaption %
    \label{fig:AO_R}
  \end{subfigure}
  \caption{Fractions of precise numbers in NA-UR, A-UR, and AO-UR}
  \label{fig:minimum_unrestr}
\end{figure}

\begin{figure}
  \begin{subfigure}{0.3\textwidth}
    \centering
    \includegraphics[width=\linewidth]{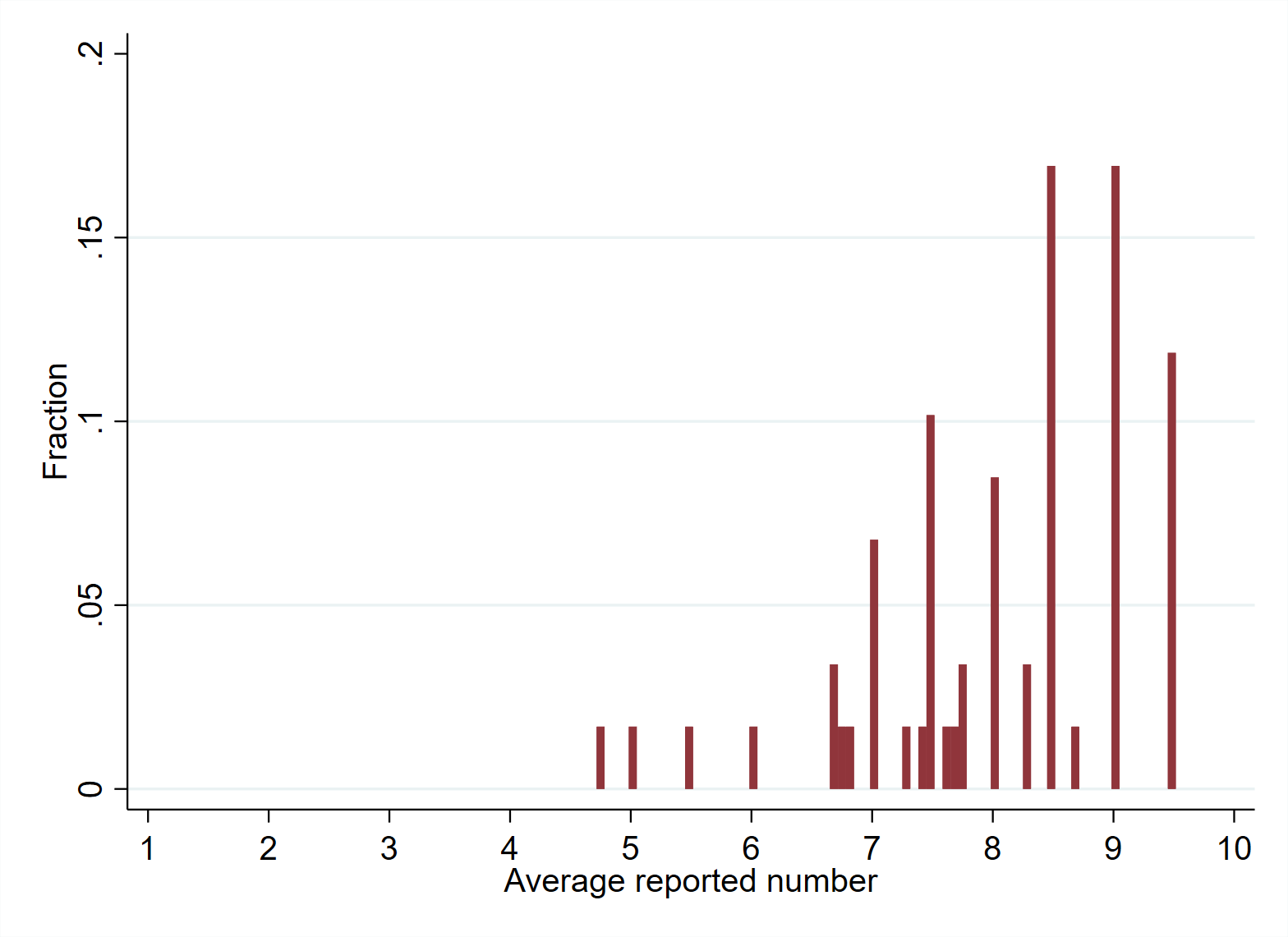}
    \phantomsubcaption %
    \label{fig:NA_R}
  \end{subfigure}%
  \hfill
  \begin{subfigure}{0.3\textwidth}
    \centering
    \includegraphics[width=\linewidth]{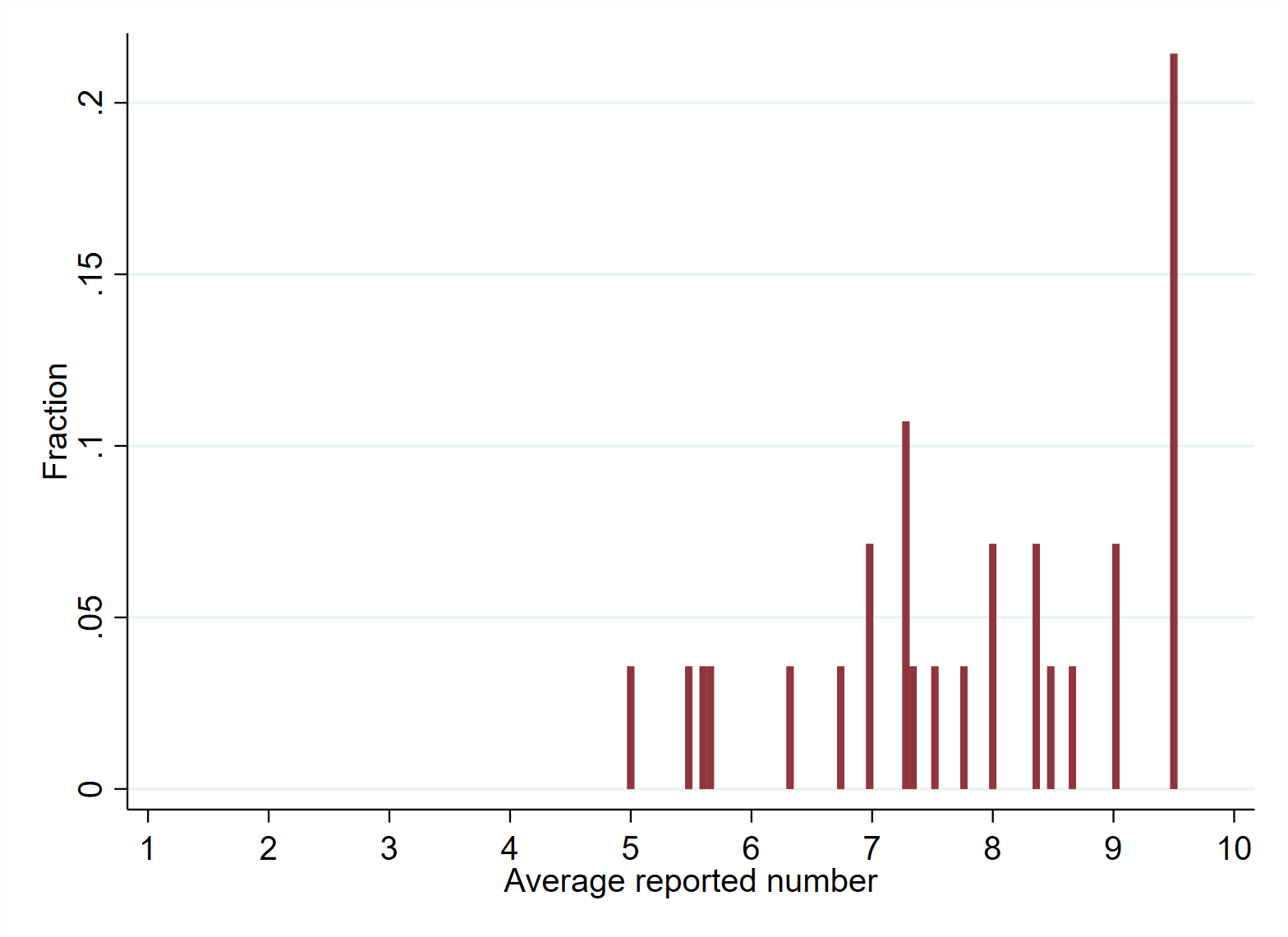}
    \phantomsubcaption %
    \label{fig:A_R}
  \end{subfigure}%
  \hfill
  \begin{subfigure}{0.3\textwidth}
    \centering
    \includegraphics[width=\linewidth]{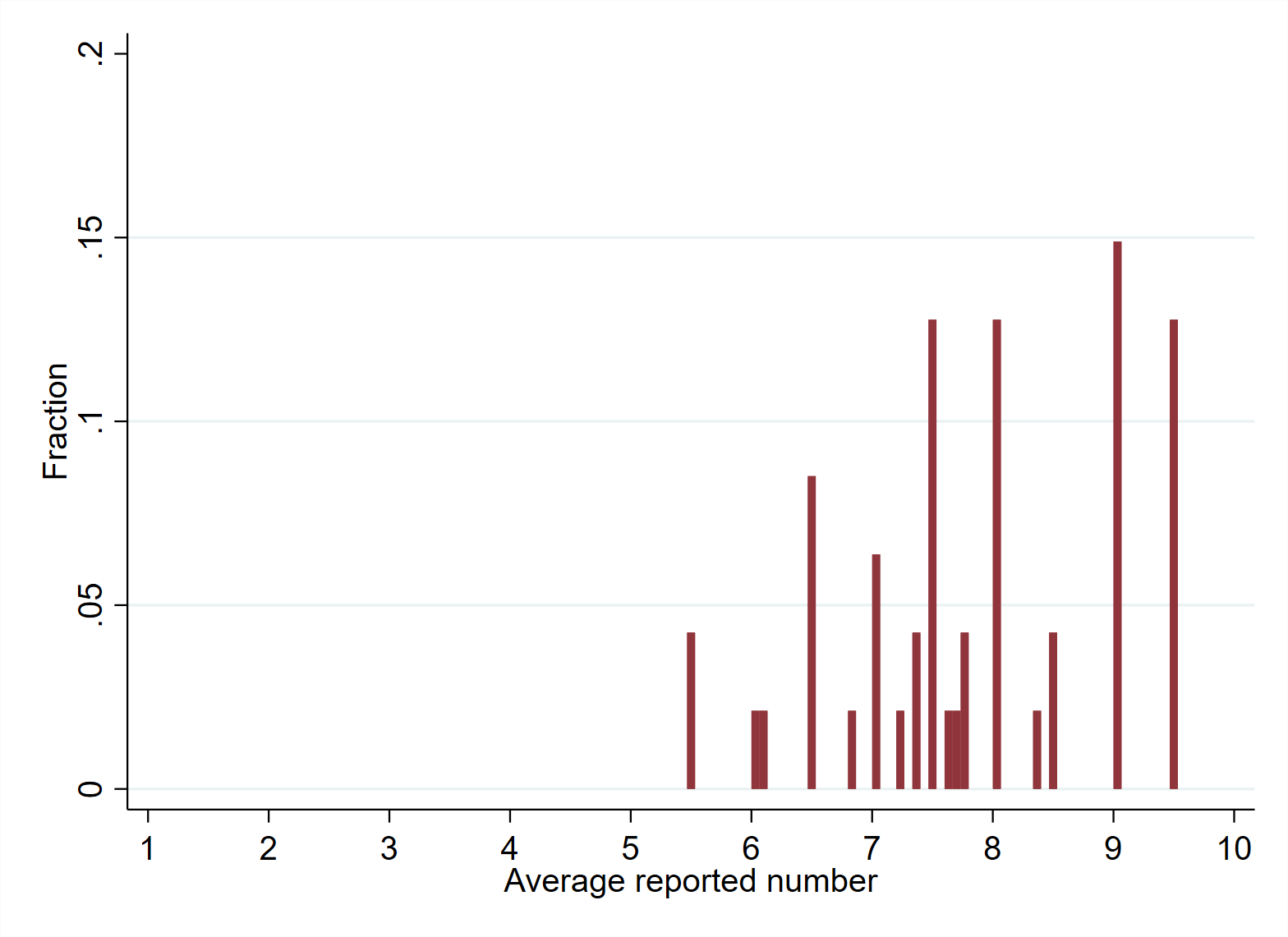}
    \phantomsubcaption %
    \label{fig:AO_R}
  \end{subfigure}
  \caption{Fractions of average vague messages in NA-UR, A-UR, and AO-UR}
  \label{fig:minimum_unrestr}
\end{figure}

\begin{figure}
  \begin{subfigure}{0.3\textwidth}
    \centering
    \includegraphics[width=\linewidth]{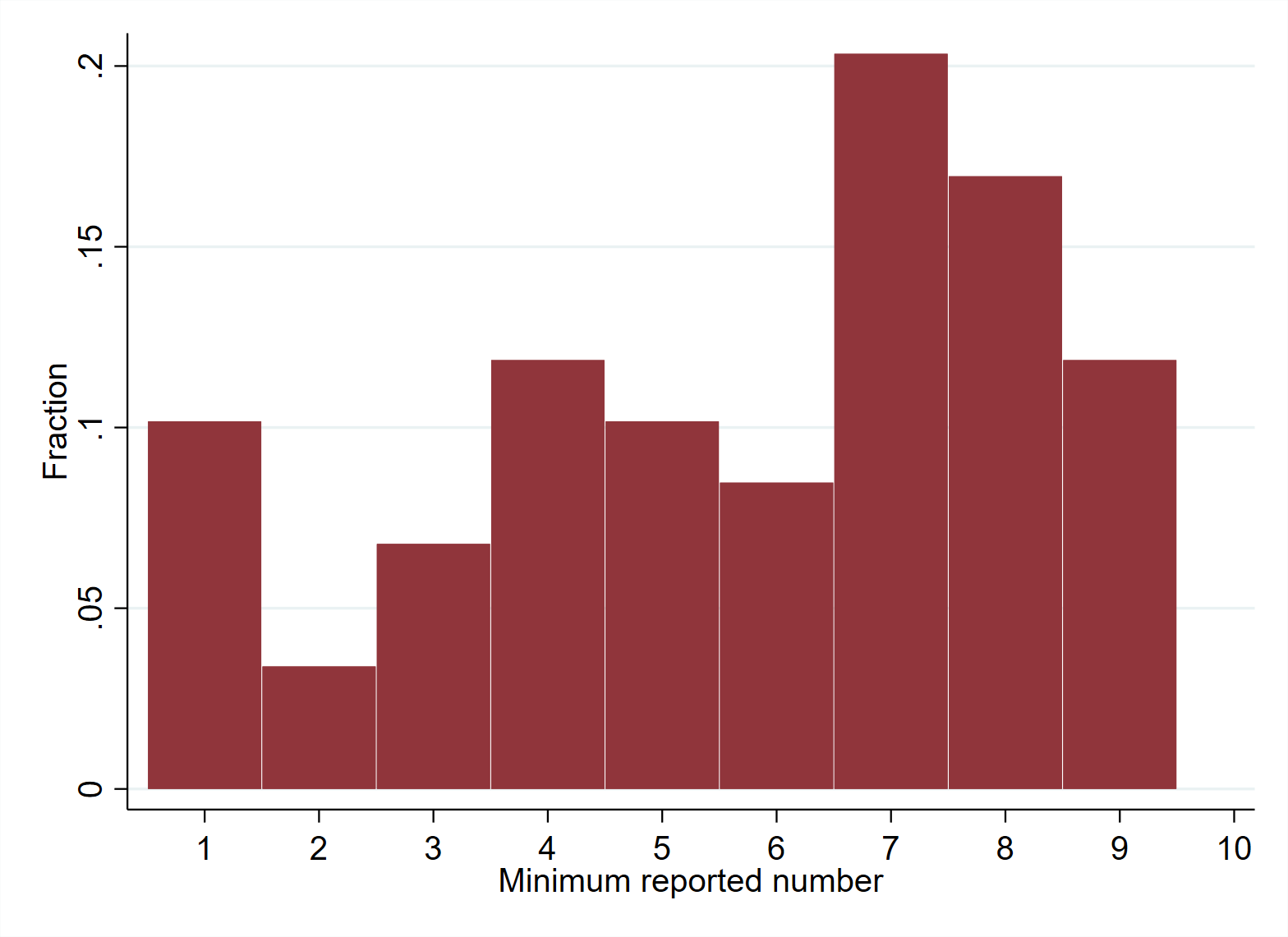}
    \phantomsubcaption %
    \label{fig:NA_R}
  \end{subfigure}%
  \hfill
  \begin{subfigure}{0.3\textwidth}
    \centering
    \includegraphics[width=\linewidth]{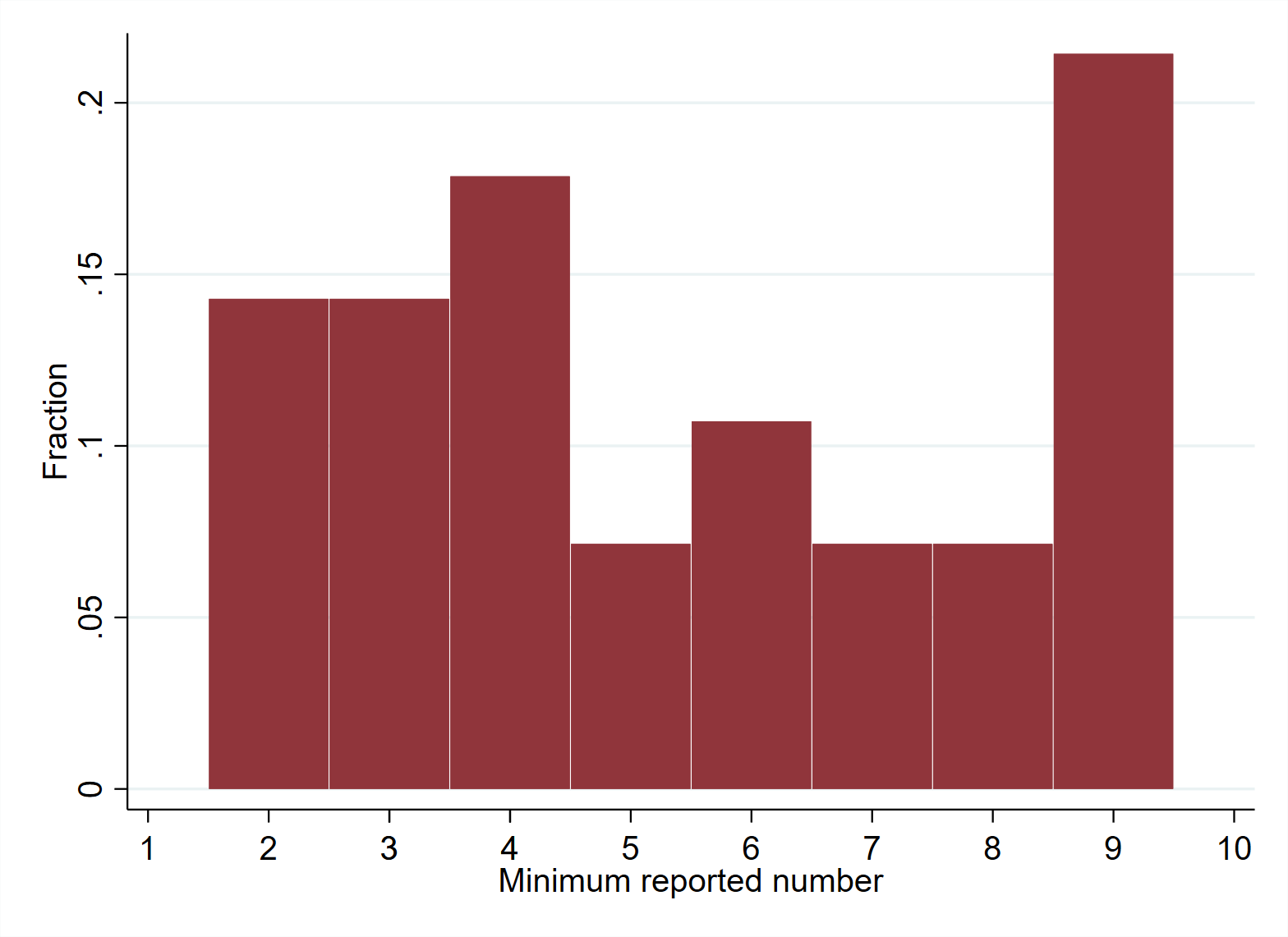}
    \phantomsubcaption %
    \label{fig:A_R}
  \end{subfigure}%
  \hfill
  \begin{subfigure}{0.3\textwidth}
    \centering
    \includegraphics[width=\linewidth]{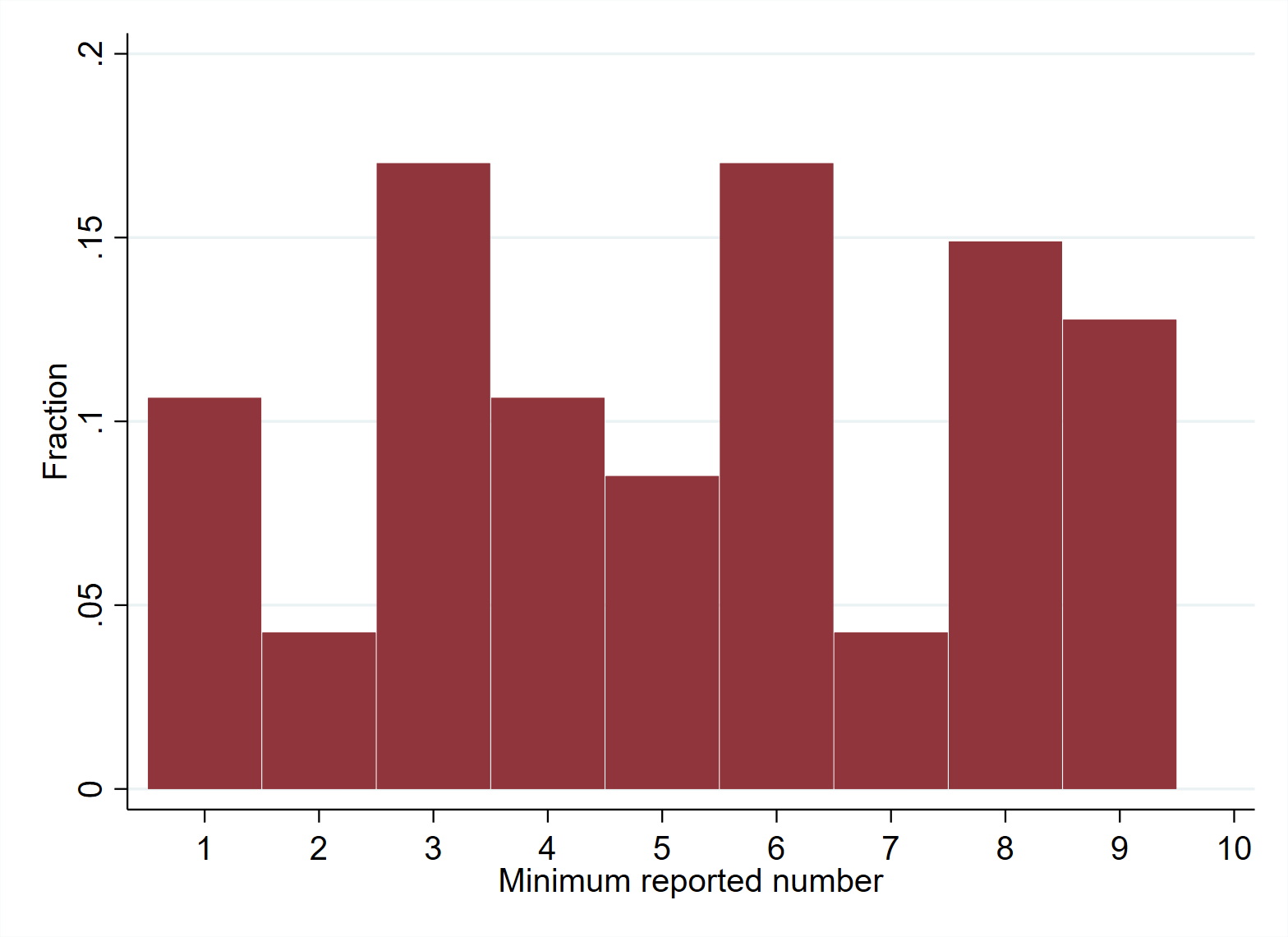}
    \phantomsubcaption %
    \label{fig:AO_R}
  \end{subfigure}
  \caption{Fractions of minimum reported vague messages in NA-UR, A-UR, and AO-UR}
  \label{fig:minimuM_vague_unrestr}
\end{figure}

\end{document}